\documentclass[a4paper,11pt]{article}

\usepackage{geometry}										% [showframe]
\usepackage[english]{babel}									% language
\usepackage[utf8]{inputenc}									% input text encoding
\usepackage[T1]{fontenc}									% output encoding

\usepackage{lmodern}										% Latin Modern

\usepackage{amsmath}										% improved mathematical typesetting
\usepackage{amstext}										% text in mathematics mode
\usepackage{amsfonts}										% extended set of fonts
\usepackage{amssymb}										% extended symbol collection

\usepackage{bm}											% \bm{} for bold math (greek) symbols, alternatives: \mathbf{}, \boldsymbol{}, \pmb{}
\usepackage{dsfont}										% \mathds{N}, \mathds{R}, \mathds{1}
\usepackage{units}										% \unit[1]{mm}, \unit[1]{ly}

\usepackage{textcomp}										% \textcopyright
\usepackage{hyphenat}                                       % hyphen

\usepackage{graphicx}										% graphics package
\usepackage[svgnames]{xcolor}									% advanced version of colors
\usepackage{booktabs}										% nice tables
\usepackage{multirow}										% cells that span multiple rows
\usepackage[inline]{enumitem}
\usepackage{algorithm}
\usepackage{algorithmic}
\usepackage{appendix}

\graphicspath{{./fig/}}

\usepackage{mathtools}

\newcommand{\mathbbm}[1]{\text{\usefont{U}{bbm}{m}{n}#1}} 
\newcommand{\Esp}[1]{{\mathbb E}\left[ #1 \right]}

\newcommand{\Var}[1]{{\rm Var}\left[ #1 \right]}
\newcommand{\Vare}[2]{{\rm Var}_{#1}\left[#2\right]}
\newcommand{\Cov}[1]{{\rm Cov}\left[ #1 \right]}
\newcommand{\ve}[1]{\boldsymbol{#1}}
\newcommand{\acc}[1]{\left\{#1\right\}}
\newcommand{\eqdef}{\stackrel{\text{def}}{=}}

\DeclarePairedDelimiter\abs{\lvert}{\rvert}
\DeclarePairedDelimiter\norm{\lVert}{\rVert}
\makeatletter
\let\oldabs\abs
\def\abs{\@ifstar{\oldabs}{\oldabs*}}
\let\oldnorm\norm
\def\norm{\@ifstar{\oldnorm}{\oldnorm*}}
\makeatother

\newcommand{\Ve}[1]{\boldsymbol{#1}}

\newcommand{\ca}{{\mathcal A}}
\newcommand{\cc}{{\mathcal C}}
\newcommand{\cd}{{\mathcal D}}

\newcommand{\ch}{{\mathcal H}}
\newcommand{\cl}{{\mathcal L}}
\newcommand{\cm}{{\mathcal M}}
\newcommand{\cn}{{\mathcal N}}
\newcommand{\cu}{{\mathcal U}}
\newcommand{\cx}{{\mathcal X}}
\newcommand{\cy}{{\mathcal Y}}
\newcommand{\Rr}{{\mathbb R}}
\newcommand{\Nn}{{\mathbb N}}

\newcommand{\GLD}{{\rm GLD}}
\newcommand{\GLaM}{{\rm GLaM}}
\newcommand{\tj}{{\rm traj}}
\newcommand{\MC}{{\rm MC}}

\newcommand{\Exp}{{\rm Exp}}
\newcommand{\SNR}{{\rm SNR}}
\newcommand{\QoI}{{\rm QoI}}

\DeclareMathOperator*{\PC}{PC}

\DeclareMathOperator{\Betafun}{B}

\DeclareMathOperator{\AOLS}{AOLS}

\DeclareMathOperator{\WLS}{WLS}

\newcommand{\D}{\mathrm{d}}

\newcommand{\iu}{{\boldsymbol{\mathsf{u}}}}

\providecommand{\ie}{i.e.,\;}
\providecommand{\eg}{e.g.,\;}

\newlength{\HYDROsubWidth}	
\newlength{\HYDROfigHeight}	
\newlength{\HYDROmapHeight}	
\newlength{\HYDROfigHeightNew}
\usepackage{authblk}										% [auth-sc,affil-it], to be loaded after titling
\title{Global sensitivity analysis for stochastic simulators based on
	generalized lambda surrogate models}
\author[1]{Xujia Zhu \thanks{zhu@ibk.baug.ethz.ch}}
\author[1]{Bruno Sudret\thanks{sudret@ethz.ch}}
\affil[1]{Chair of Risk, Safety and Uncertainty Quantification, ETH Z\"{u}rich, Stefano-Franscini-Platz 5, 8093 Z\"{u}rich, Switzerland}

\date{\today}

\usepackage{microtype}										% typographical refinements

\usepackage{indentfirst}									% indents first paragraph in new section

\usepackage{caption}
\usepackage{subcaption}

\usepackage[numbers,sort&compress]{natbib}							% number ranges are compressed and hyphenated
\usepackage{hyperref}
\hypersetup{
	pdftitle = {Global sensitivity analysis for stochastic simulators based on 
		generalized lambda surrogate models},
	pdfauthor = {Xujia Zhu, Bruno Sudret},
	pdfsubject = {Manuscript submitted to Reliab. Eng. Sys. Safety},
	pdfkeywords = {Stochastic simulators, Surrogate modeling, Sensitivity analysis, 
		Sobol' indices, Generalized lambda distributions, Polynomial chaos expansions},
	colorlinks = false,										% false: boxed links; true: colored links
}

\usepackage[capitalize]{cleveref}								% Eq., Fig., Table, Section, has to be loaded after hyperref
\creflabelformat{equation}{#2(#1)#3}								% hyperlinks covers the parenthesis around equation
\crefrangelabelformat{equation}{#3(#1)#4 to #5(#2)#6}						% multiple references
\labelcrefformat{subequation}{#2(#1)#3}								% the same for subequations
\labelcrefrangeformat{subequation}{#3(#1)#4 to #5(#2)#6}					% mutiple references

\begin{document}

\maketitle

\begin{abstract}
Global sensitivity analysis aims at quantifying the impact of input variability 
onto the variation of the response of a computational model. It has been widely 
applied to deterministic simulators, for which a set of input parameters has a 
unique corresponding output value. Stochastic simulators, however, have 
intrinsic 
randomness due to their use of (pseudo)random numbers, so they give different 
results when run twice with the same input parameters but non-common random 
numbers. Due to this random nature, conventional Sobol' indices, used 
in global sensitivity analysis, can be 
extended to stochastic simulators in different ways. In this paper, we discuss 
three possible extensions and focus on those that depend 
only on the statistical dependence between input and output. This 
choice ignores the detailed data generating process involving the internal 
randomness, and can thus be applied to a wider class of problems. We propose to 
use the generalized lambda model to emulate the response distribution of 
stochastic simulators. Such a surrogate can be constructed without the need for replications. The proposed method is applied to three examples including two case studies in finance and epidemiology. The results confirm the convergence of the approach for estimating the sensitivity indices even with the presence of strong heteroskedasticity and small signal-to-noise ratio.
\end{abstract}

\section{Introduction}
\label{sec:intro}

Computational models, a.k.a. simulators, have been extensively used to 
represent physical phenomena and engineering systems. They can help
assess the reliability, control the risk and optimize the behavior of 
complex systems early at the design stage. Conventional simulators are usually 
deterministic, in the sense that repeated model evaluations with the same input 
parameters yield the same value of the output. In contrast, several runs of a 
stochastic simulator for a given set of 
input parameters provide different results. More precisely, the output of a 
stochastic simulator is a random variable following an unknown probability 
distribution. Hence, each model evaluation with 
the same input values generates a realization of the random variable. 
Mathematically, a stochastic simulator can be defined by 
\begin{equation}\label{eq:defsto}
\begin{split}
\cm_s: \cd_{\ve{X}} \times \Omega &\rightarrow \Rr \\
(\ve{x},\omega) &\mapsto \cm_s(\ve{x},\ve{Z}(\omega)) ,
\end{split}
\end{equation}
where $\ve{x}$ is the input vector that belongs to the input space 
$\cd_{\ve{X}}$, and $\Omega$ denotes 
the probability space that represents the stochasticity. The intrinsic 
randomness is due to the fact that some \emph{latent variables} 
$\ve{Z}(\omega)$ inside the model are not explicitly considered as a part 
of the input variables: given a fixed input value $\ve{x}_0$, the output of the simulator is a random variable. 

In this respect, we can consider a stochastic simulator as a random field 
indexed by the parameters $\ve{x} \in \cd_{\ve{X}}$ \citep{AzziIJUQ2019}. For a 
given realization of the latent variables $\ve{z}_0$, the simulator becomes a 
deterministic function of $\ve{x}$. This is realized in practice by initializing the random seed to the same value before running the simulator for different $\ve{x}$' s, a trick known as \emph{common random numbers}. The (classical) functions $\ve{x} \mapsto \cm_s(\ve{x}, \ve{z}_0)$ will be called \emph{trajectories} in this paper. One particular trajectory corresponds to one particular value $\ve{z}_0$. 

In contrast, for a given $\ve{x}_0 \in \cd_{\ve{X}}$, the output of the 
stochastic simulator is a random variable. Its distribution can be obtained by 
repeatedly running the simulator with $\ve{x}_0$, yet different realizations of 
the latent variables called \emph{replications}.

Stochastic simulators are ubiquitous in modern engineering, finance and medical 
sciences. Typical examples include stochastic 
differential equations (\eg financial models \cite{McNeil2005}) and 
agent-based models (\eg epidemiological models \cite{Britton2010}). 
To a certain extent, physical experiments can also be considered as stochastic 
models, because we may not be able to measure and consider all the relevant 
variables that can uniquely determine the experimental conditions.

In practice, the input variables may be affected by uncertainty due to noisy 
measurements, expert judgment or lack of knowledge. Therefore, they are 
modeled as random variables and grouped into a random vector 
$\ve{X}=\left(X_1,X_2,\ldots,X_M\right)$, which is 
characterized by a joint distribution $f_{\ve{X}}$. Quantification of the 
contribution of input variability to the output uncertainty is a major 
task in sensitivity analysis \cite{Saltelli2000}. It allows us to identify the 
most important set of input variables that dominate the output variability and 
also to figure out non-influential variables. This information provides more 
insights into the simulator and can be further used for model calibrations and 
decision making \cite{Rocquigny2008}. 

A large number of methods have been successfully developed to perform 
sensitivity analysis in the context of deterministic simulators 
\cite{Saltelli2000,Helton2006,Borgonovo2017}. Among others, the variance-based 
sensitivity analysis, also referred to as Sobol' indices \cite{Sobol1993}, is 
one of the most popular approaches, which relies on the analysis of variance. 
Several extensions of Sobol' indices to stochastic simulators can be found in 
the literature, depending on the treatment of the intrinsic randomness. 
It is worth emphasizing that the overall uncertainty now consists of two parts, 
namely the inherent stochasticity in the latent variables and the uncertainty 
in the input parameters $\ve{X}$.
Iooss and Ribatet \cite{Iooss2009} include the latent variables as a part of 
the input, which results in a natural extension of the classical Sobol' indices 
to stochastic simulators. Hart et al. \cite{Hart2016} and Jimenez et al. 
\cite{Jimenez2017} define the Sobol' indices as functions of the 
latent variables which, as a consequence, become random variables whose 
statistical properties can be studied. 
Recently, Azzi et al. \cite{AzziSens2020} propose to represent the intrinsic 
randomness by the entropy of the response distribution and to calculate the  
classical Sobol' indices on the latter. All in all, relatively little 
attention has been devoted to sensitivity analysis for stochastic simulators.

Sensitivity analysis usually requires a large number of model evaluations for 
different realizations of the input vector. Due to the intrinsic randomness
of stochastic simulators, an additional layer of stochasticity comes 
on top of the input uncertainty, which requires repeated runs with the
same input parameters to fully characterize the 
model response. As a consequence, such analyses become intractable when the 
simulator is expensive to evaluate. To alleviate the computational burden, 
surrogate models can be constructed to mimic the original numerical model at a 
smaller computational cost. Large efforts have been dedicated to emulating the 
mean and variance function of stochastic simulators 
\cite{Dacidian1987,Marrel2012,Binois2018}. These two functions provide only
the first two moments of the response distribution and are mostly used 
to estimate the Sobol' indices proposed in \cite{Iooss2009}. In recent papers 
\cite{ZhuIJUQ2020,ZhuSIAMUQ2021}, we developed a novel surrogate model, called 
\textit{generalized lambda model} (GLaM), to emulate the whole  
response 
distribution of stochastic simulators. This model uses generalized lambda 
distributions (GLD) \cite{Karian2000} to flexibly approximate the response 
distribution, 
while the distribution parameters cast as functions of the inputs 
are approximated through polynomial chaos expansions (PCE) 
\cite{Ghanembook2003}.

Based on these premises, the goal of this paper is to establish a clear 
framework to carry out global sensitivity analysis for stochastic simulators, 
and to propose efficient computational approaches based on GLaM stochastic 
emulators. Therefore, the original contributions of this paper are two-fold. 
On the one hand, we give a thorough review of the current development of global 
sensitivity analysis for stochastic simulators. We point out the nature and 
the properties of different extensions of Sobol' indices, which provides a general 
guideline to their usage. On the other hand, we present a unified framework 
based on generalized lambda models to calculate a whole variety of global 
sensitivity indices using this single surrogate.

The paper is organized as follows. First, we review three 
extensions of Sobol' indices to stochastic simulators in \Cref{sec:GA4SS}. In 
\Cref{sec:GLaMs}, we present the framework of GLaMs \cite{ZhuIJUQ2020}. There, 
we recap the fitting procedure proposed in \cite{ZhuSIAMUQ2021}, where it is 
emphasized that there is no need for replicated runs. Then, we discuss the use 
of GLaMs for estimating different types of Sobol' indices. In 
\Cref{sec:examples}, we illustrate the performance of GLaMs on three examples. 
While the first example is analytical, the second and third ones are realistic 
case studies in finance and epidemiology, respectively. Finally, we summarize 
the main findings of the paper and provide an outlook for future research in 
\Cref{sec:conclusions}.

\section{Global sensitivity analysis of stochastic simulators}
\label{sec:GA4SS}

\subsection{Sobol' indices}
\label{sec:Sobol}
Variance-based sensitivity analysis has been extensively studied and 
successfully developed in the context of deterministic simulators. For a 
deterministic model $\cm_d$, Sobol' indices quantify the contribution of each 
input variable $\acc{X_i, i=1,\ldots,M}$, or combination thereof, to the 
variance of the model output $Y=\cm_d(\ve{X})$. 

In this paper, we assume that $X_i$'s are mutually independent. Let us split 
the input vector into two subsets $\ve{X} = 
(\ve{X}_{\iu},\ve{X}_{\sim\iu})$, where $\iu \subset \acc{1,\ldots,M}$ and 
$\sim \iu$ is the complement of $\iu$, \ie $\sim 
\iu = \acc{1,\ldots,M} \setminus \iu$. From the total variance theorem, the 
variance of the output can be decomposed as
\begin{equation}\label{eq:vardecomp}
\Var{Y} = \Esp{\Var{Y \mid \ve{X}_{\iu}} } + \Var{\Esp{Y \mid \ve{X}_{\iu}} 
}.
\end{equation}
The first-order and total Sobol' indices introduced by Sobol'
\cite{Sobol1993} 
and Homma and Saltelli \cite{Homma1996} for the subset of input variables 
$\ve{X}_{\iu}$ are defined by 
\begin{equation}\label{eq:DeterSobol}
S_{\iu} \eqdef \frac{\Var{\Esp{Y\mid\ve{X_\iu}}}} {\Var{Y}}, \quad	S_{T_\iu} 
\eqdef 1 - \frac{\Var{\Esp{Y\mid\ve{X_{\sim\iu}}}}}{\Var{Y}} = 1 - S_{\sim \iu}.
\end{equation}
Higher-order Sobol' indices can be defined with the help 
of $S_{\iu}$. For example, the second-order or two-factor interaction
Sobol' index of $X_1$ and $X_2$ is given by
\begin{equation}
S_{1,2} \eqdef S_{\acc{1,2}} - S_1 - S_2,
\end{equation}
where we denote $S_{\acc{i}}$ by $S_i$ for the sake of simplicity.

In the context of stochastic simulators defined in \Cref{eq:defsto}, the input 
variables alone do not determine the value of the output. Iooss and 
Ribatet \cite{Iooss2009} extend $\ve{X}$ by adding the internal source of 
randomness represented by latent variables $\ve{Z}$, which turns the 
stochastic simulator into a deterministic one. In this case, 
all the input variables are gathered in $\left(\ve{X}_{\iu},\ve{X}_{\sim 
\iu},\ve{Z}\right)$, and thus the Sobol' indices in \Cref{eq:DeterSobol} can be 
naturally extended to
\begin{equation}\label{eq:ClassSobol}
S_{\iu} \eqdef \frac{\Var{\Esp{Y\mid\ve{X_\iu}}}} {\Var{Y}}, \quad	
S_{T_\iu} 
\eqdef 1 - \frac{\Var{\Esp{Y\mid\ve{X_{\sim\iu}},\ve{Z} }}}{\Var{Y}}.
\end{equation}
Note that $S_{\iu}$ has the same expression as in the case of deterministic 
simulators, but $S_{T_\iu}$ contains the additional variables $\ve{Z}$ 
\cite{Marrel2012}. Similarly, higher-order Sobol' indices corresponding to 
interactions among components of $\ve{X}$ are defined in the same way as 
deterministic simulators, whereas interactions between components of $\ve{X}$ 
and $\ve{Z}$ involve $\ve{Z}$ in their definition. Since this is a direct 
extension, the Sobol' indices defined in \Cref{eq:ClassSobol} are referred to 
as \emph{classical Sobol' indices} in the sequel.

Another way to extend Sobol' indices to stochastic simulators is to first 
eliminate the internal randomness by representing the response 
random variable $Y(\ve{x})$ by some summarizing statistical quantity, called 
here quantity of interest (QoI) denoted by $\QoI(\ve{x})$, such as the 
mean value $m(\ve{x})$, 
variance $v(\ve{x})$ \cite{Iooss2009}, $\alpha$-quantile $q_{\alpha}(\ve{x})$ 
\cite{Browne2016} and differential entropy $h(\ve{x})$ \cite{AzziSens2020}. As 
a result, the stochastic simulator is reduced to a deterministic function 
$\QoI(\ve{x})$, and we can calculate the associated 
\emph{QoI-based	Sobol' indices} as follows:
\begin{equation}\label{eq:qSobol}
S^{\QoI}_{\iu} \eqdef \frac{\Var{\Esp{\QoI(\Ve{X} )|\ve{X_\iu}}}} 
{\Var{\QoI(\Ve{X})}},
\quad
S^{\QoI}_{T_\iu} \eqdef 1 -
\frac{\Var{\Esp{\QoI(\Ve{X})|\Ve{X}_{\sim\iu}}}}{\Var{\QoI(\Ve{X})}}.% = 
%1-S^q_{\sim 
%\iu}.
\end{equation}

A third extension is defined by considering a stochastic 
simulator as a random field. For a fixed internal randomness $\ve{Z}(\omega) = 
\ve{z}$, the stochastic simulator is a deterministic function of the input 
variables, which corresponds to a trajectory. Hence, the associated Sobol' 
indices are well-defined. Yet, they are random variables because of their 
dependence on $\ve{Z}$ \cite{Hart2016}, which results in the 
\emph{trajectory-based Sobol' 
indices}:
\begin{equation}\label{eq:tdSobol}
	S^{\tj}_{\iu}(\ve{Z}) \eqdef \frac{\Vare{\ve{X}_{\iu}}{\Esp{Y \mid
				\ve{X}_{\iu},\ve{Z}}} }{\Var{Y \mid \ve{Z}}}, \quad
	S^{\tj}_{T_\iu}(\ve{Z}) \eqdef 1 - \frac{\Vare{\ve{X}_{\sim\iu}}{Y 
			\mid \ve{X}_{\sim\iu},\ve{Z}}}{\Var{Y \mid \ve{Z}}},
\end{equation}
where the indices of the variance operators correspond 
to those variables to which these operators apply.

\subsection{Discussion}
\label{sub:discussions}
The three types of Sobol' indices introduced above have different nature and 
focus. The classical Sobol' indices defined in \Cref{eq:ClassSobol} treat the 
latent variables $\ve{Z}$ as a set of separate input variables. As a 
result, indices of this type treat $\ve{Z}$ in the same way as $\ve{X}$. 
The first-order index $S_{\iu}$ indicates how much the output variance can 
be reduced (in expectation) if we can fix the value of $\ve{X}_{\iu}$. Besides, 
the classical Sobol' indices can also quantify the influence of the intrinsic 
randomness as well as its interactions with input variables. 

The QoI-based Sobol' indices defined in \Cref{eq:qSobol} help study a 
specific statistical quantity of the model response, which is a deterministic 
function of the inputs. Using a summary quantity to represent the random output might lead to a loss of information \cite{Hart2016}, unless this quantity itself is of interest. For example, we may want to find the variable(s) that has the largest effect(s) on the 95\% quantile (i.e., $q_{\alpha}(\ve{X})$ for $\alpha = 0.95$) of the model response. However, the importance (ranking) of the inputs $X_i$'s can be quite different depending on the choice of the QoI. 

Unlike the previous two types of indices, trajectory-based Sobol' indices presented in \Cref{eq:tdSobol} are random variables. This is because the latent variables $\ve{Z}$ and the input parameters $\ve{X}$ are treated differently: conditioned on a given $\ve{Z}=\ve{z}_0$, the stochastic simulator reduces to a deterministic function of $\ve{X}$, and we can calculate the associated (classical) Sobol' indices. To evaluate the probability distribution of these trajectory-based Sobol' indices requires that the same random seeds can be explicitly fixed in the simulator when running it for different values of $\ve{x}$. In a sense, trajectory-based Sobol' indices emphasize the variation of the trajectory of stochastic simulators. They typically show the importance of each input variable in terms of its contribution to the variability of trajectories. 

For stochastic simulators, the question ``which input variable has the strongest effect?'' is rather vague and cannot be answered by a single type of index. The analyst should properly pose the problem and select the appropriate sensitivity measures. If one aims at reducing the variance of the model output, the classical Sobol' indices are of interest. If one is interested in 
some summary QoIs (which is often the case for applications, e.g., quantiles in reliability analysis), the QoI-based indices are more appropriate. Finally, if one can control the internal randomness and is interested in the variability of the model output for fixed intrinsic stochasticity, trajectory-based indices should be selected. 

To further illustrate this discussion, let's consider the stochastic simulator
\begin{equation*}
	Y(\ve{X}) = \cm_s(\ve{X},Z) = X_1 + X_2\cdot Z
\end{equation*}
where $X_1$, $X_2$ and $Z$ are independent random variables following standard normal distributions. The classical first-order Sobol' indices are $S_1 = 0.5$ and $S_2 = 0$. Therefore, if we want to primarily reduce the variance of $Y$, $X_1$ should be investigated. For the response mean function $m(\ve{X}) = X_1$, we have $S^m_1 = 1$ and $S^m_2 = 0$, which indicates that $X_1$ contributes fully to the variation of the mean function. In contrast, for the variance function $v(\ve{X}) = X^2_2$, $S^v_1 = 0$ and $S^v_2 = 1$ reveals a different order. Regarding the trajectory-based indices, we have $S^{\tj}_1(Z) = \frac{1}{1+Z^2}$ and $S^{\tj}_2(Z) = \frac{Z^2}{1+Z^2}$ which are two random variables (due to the randomness in $Z$). The probability distribution of the two variables characterize how the randomness in $\ve{X}$ affects the function $\cm_s(\ve{X},z)$ with $z$ being fixed as a single realization of $Z$. In summary, even for this simple toy example, the various indices provide different conclusions.

It is worth remarking that the classical Sobol' indices $S_{\iu}$ in 
\Cref{eq:ClassSobol} share some common properties with the mean-based Sobol' 
indices in \cref{eq:qSobol}, when we consider the mean function $\QoI(\ve{x}) 
\eqdef m(\ve{x}) = \Esp{Y \mid \ve{X} = \ve{x}}$:
\begin{equation}\label{eq:mSobol}
S^{m}_{\iu} = \frac{\Var{\Esp{m(\Ve{X})|\ve{X_\iu}}}}{\Var{m(\Ve{X})}}.
\end{equation} 
According to the law of total expectation, we have 
\begin{equation}\label{eq:lttp}
\Esp{Y \mid \ve{X}_{\iu}} = \Esp{\Esp{Y \mid \ve{X}}\mid\ve{X}_{\iu}} = 
\Esp{m(\ve{X})\mid \ve{X}_{\iu}}.
\end{equation}
As a result, $S_{\iu}$ can be rewritten as
\begin{equation}\label{eq:fCSobol}
S_{\iu} = \frac{\Var{\Esp{Y\mid\ve{X_\iu}}}} {\Var{Y}} = 
\frac{\Var{\Esp{m(\ve{X})\mid\ve{X_\iu}}}} {\Var{Y}},
\end{equation}
This implies that both classical Sobol' indices $S_{\iu}$ and mean-based Sobol' 
indices $S^m_{\iu}$ provide the same ranking, as the numerators of 
\Cref{eq:mSobol,eq:fCSobol} are identical. However, it is worth emphasizing 
that $S_{\iu}$ is not equal to $S^m_{\iu}$ and they are measuring 
different quantities, since the denominator of 
\Cref{eq:fCSobol} is $\Var{Y}$ but that of \Cref{eq:mSobol} is 
$\Var{m(\ve{X})}$. 

As a summary for all three extensions, a stochastic simulator is essentially 
transformed into a reduced deterministic model at a certain stage. The 
classical Sobol' indices include the latent variables as a part 
of the inputs. The QoI-based Sobol' indices rely on a 
deterministic representation. The trajectory-based 
indices are random variables whose statistical properties can be studied at the 
cost of repeating a standard Sobol' analysis for different realizations of the 
latent variables separately. As a result, the three types of 
sensitivity indices can be estimated by modifying only slightly the standard 
methods based on Monte Carlo simulation developed for deterministic simulators 
\cite{Saltelli2000}. 

The classical Sobol' indices involving $\ve{Z}$ (\eg $S_{\ve{Z}}$, 
$S_{T_{\iu}}$ in \cref{eq:ClassSobol}) and the trajectory-based Sobol' 
indices require controlling the latent variables $\ve{Z}$. In practical 
computations, this is achieved by fixing the \emph{random seed} in the 
computational model \cite{Marrel2012,Hart2016}. However, for certain types of 
stochastic simulators, or when the data are generated by physical experiments,
it may be difficult to control or even identify $\ve{Z}$. For the sake of general applicability, we focus in this paper only on Sobol' indices that can be estimated by manipulating $\ve{X}$, that is, the QoI-based Sobol' indices and, to some extent, the classical Sobol' indices. 

Using Monte Carlo simulations to estimate these indices requires evaluating the 
simulator for various realizations of the input vector. In addition, it is 
generally necessary to evaluate the function $\QoI(\ve{x})$ for calculating the 
associated QoI-based Sobol' indices. However, this function is not directly accessible due to the intrinsic randomness. Therefore this function is usually estimated by using replications; i.e., for each realization $\ve{x}$, the simulator is run many times, and $\QoI(\ve{x})$ is estimated from the output samples. Both factors call for a large number 
of model runs, which becomes impracticable for costly models. Therefore, the use of surrogate models is unavoidable.

In the sequel, we present the 
generalized lambda model as a stochastic surrogate. Such a model emulates the 
response distribution conditioned on $\ve{X} = \ve{x}$, which fully 
characterizes the statistical dependence between the inputs and output. 
Therefore, it can be used to estimate the 
considered Sobol' indices.

\section{Generalized lambda models}
\label{sec:GLaMs}
Generalized lambda models consist of mainly two parts: the generalized lambda 
distribution and polynomial chaos expansions. In this section, we briefly recap 
these two elements and present an algorithm to construct such a model without 
the need for replicated runs of the simulator. Then, we discuss how to estimate 
the sensitivity indices from the surrogate.

\subsection{Generalized lambda distributions}
\label{sec:GLD}
The generalized lambda distribution is a flexible distribution family, 
which is designed to approximate many common distributions 
\cite{Karian2000}, \eg normal, lognormal, Weibull and generalized extreme value 
distributions. A GLD is defined by its \emph{quantile function} $Q(u)$ with $u 
\in [0,1]$, that is, the inverse of the cumulative distribution function $Q(u) 
= F^{-1}(u)$. In this paper, we consider the GLD of the 
Freimer-Kollia-Mudholkar-Lin (FKML) family \cite{Freimer1988} with four 
parameters, whose quantile function is defined by
\begin{equation}\label{eq:FKML}
Q(u;\ve{\lambda}) = \lambda_1 + \frac{1}{\lambda_2}\left( 
\frac{u^{\lambda_3}-1}{\lambda_3} 
- \frac{(1-u)^{\lambda_4}-1}{\lambda_4}\right),
\end{equation}
where $\lambda_1$ is the location parameter, $\lambda_2$ is the scaling 
parameter, and $\lambda_3$ and $\lambda_4$ are the shape parameters. 
$\lambda_2$ is required to be positive to produce valid quantile functions (\ie 
$Q$ being non-decreasing on $[0,1]$). Based on the quantile function defined in 
\Cref{eq:FKML}, the probability density function (PDF) of a random 
variable $Y$ following a GLD is given by
\begin{equation}\label{eq:FKMLpdf}
f_Y(y;\ve{\lambda}) = \frac{f_U(u)}{Q^\prime(u;\ve{\lambda})} 
=\frac{\lambda_2}{ 
	u^{\lambda_3-1} + 
	(1-u)^{\lambda_4-1}} \mathbbm{1}_{[0,1]}(u), \text{ with } u = 
	Q^{-1}(y;\ve{\lambda}) ,
\end{equation}
where $\mathbbm{1}_{[0,1]}$ is the indicator function. A closed-form expression 
of $Q^{-1}$ is not available for arbitrary values of $\lambda_3$ and 
$\lambda_4$. Therefore, evaluating the PDF for a given $y$ usually requires 
solving the nonlinear equation (\ref{eq:FKMLpdf}) numerically. 

GLDs cover a wide range of shapes determined by $\lambda_3$ and 
$\lambda_4$ \cite{ZhuIJUQ2020}. For instance, $\lambda_3 = 
\lambda_4$ produces symmetric PDFs, and $\lambda_3 < \lambda_4$ ($\lambda_3 > 
\lambda_4$) results in left-skewed (resp. right-skewed) distributions. 
Moreover, $\lambda_3$ and $\lambda_4$ are closely linked to the support and the 
tail behaviors of the corresponding PDF. More precisely, $\lambda_3$ and 
$\lambda_4$ control the left and right tail, respectively. Whereas $\lambda_3 > 0$ implies that the PDF support is left-bounded, $\lambda_3 \leq 0$ implies that the distribution has a lower infinite support. Similarly, $\lambda_4>0$ implies that the PDF support is right-bounded, whereas it is $+\infty$ for $\lambda_4 \leq 0$. In addition, for $\lambda_3<0$ ($\lambda_4<0$), the left (resp. right) tail decays asymptotically as a power law. Hence, GLDs can also provide 
fat-tailed distributions. The reader is referred to 
\cite{ZhuIJUQ2020,Freimer1988} for a longer presentation of GLDs.

\subsection{Polynomial chaos expansions}
\label{sec:PCE}
Consider a deterministic model $\cm_d$ that maps a set of input 
parameters $\ve{x} =\left(x_1,x_2,\ldots,x_M\right) \in \cd_{\ve{X}} 
\subset\Rr^M$ to the output $y \in \Rr$. Under the 
assumption that $Y = \cm_d(\ve{X})$ has finite variance, $\cm_d$ belongs to the 
Hilbert space $\ch$ of square-integrable functions with respect to the inner 
product $ \langle u,v\rangle_{\ch} = \Esp{u(\ve{X})v(\ve{X})} = 
\int_{\cd_{\ve{X}}}
u(\ve{x})v(\ve{x})f_{\ve{X}}(\ve{x}) \D\ve{x}$. If the joint distribution 
$f_{\ve{X}}$ satisfies certain conditions \cite{Ernst2012}, the simulator 
$\cm_d$ 
admits a spectral representation in terms of orthogonal polynomials:
\begin{equation}
\label{eq:PCE}
Y = \cm_d(\ve{X}) = \sum_{\ve{\alpha}\in 
	\Nn^M}c_{\ve{\alpha}}\psi_{\ve{\alpha}}(\ve{X}),
\end{equation}
where $\psi_{\ve{\alpha}}$ is a multivariate polynomial basis function indexed 
by $\ve{\alpha} \in \Nn^M$, and $c_{\ve{\alpha}}$ denotes the associated 
coefficient. The 
orthogonal basis can be obtained by using tensor products of univariate 
polynomials, each of which is orthogonal with respect to the probability 
measure $f_{X_i}(x_i) \D x_i$ of the $i$-th variable $X_i$:
\begin{equation}\label{eq:PCE_basis}
\psi_{\ve{\alpha}}(\ve{x}) = \prod_{j=1}^{M} \phi^{(j)}_{\alpha_j}(x_j).
\end{equation}
Details about the construction of this generalized polynomial chaos expansion 
can be found in \cite{Xiu2002,SudretBookPhoon2015}.

The PCE defined in \Cref{eq:PCE} contains an infinite sum of terms. 
However, in practice, it is only feasible to use a finite series as an 
approximation. To this end, truncation schemes are adopted to 
select a set of basis functions defined by a finite 
subset $\ca\subset\Nn^M$ of multi-indices. A typical scheme is the hyperbolic 
(a.k.a. $q$-norm) truncation scheme \cite{BlatmanPEM2010} given by
\begin{equation}\label{eq:qnorm}
\ca^{p,q,M} = \acc{\ve{\alpha}\in \Nn^M : \|\ve{\alpha}\|_{q} \eqdef 
	\left(\sum_{i=1}^{M}\abs{\alpha_i}^q\right)^{\frac{1}{q}}\leq p},
\end{equation}
where $p$ is the maximum degree of polynomials, and $q \leq 1$ defines the 
quasi-norm $\|\cdot\|_q$. Note that with $q=1$, we obtain 
the full basis of total degree less than $p$, which corresponds to the 
\emph{standard truncation scheme}.

\subsection{Formulation of generalized lambda models}
\label{sec:GLaM}
Because of the flexibility of GLDs, we assume that the response random variable 
$Y(\ve{x})$ of a stochastic simulator for a given input vector $\ve{x}$ 
can be well approximated by a GLD. Hence, the 
associated distribution parameters $\ve{\lambda}$ are functions of the input 
variables:
\begin{equation} \label{eq:condGLD}
Y (\ve{x}) \sim 
\GLD\left(\lambda_1(\ve{x}),\lambda_2(\ve{x}),\lambda_3(\ve{x}),\lambda_4(\ve{x})
\right).
\end{equation}
\par
Under appropriate conditions discussed in \Cref{sec:PCE}, each component of 
$\ve{\lambda}(\ve{x})$ can be represented by a series of orthogonal 
polynomials. 
Because $\lambda_2(\ve{x})$ is required to be positive (see 
\Cref{sec:GLD}), the associated polynomial chaos representation is built on the 
natural logarithmic transform $\log\left(\lambda_2(\ve{x})\right)$. This 
results in the following approximations:
\begin{align}
\lambda_l\left(\ve{x}\right) &\approx \lambda^{\PC}_s\left(\ve{x};\ve{c}\right) 
= \sum_{\ve{\alpha}\in \ca_l} c_{l,\ve{\alpha}}\psi_{\ve{\alpha}}(\ve{x}), 
\quad l = 1,3,4, \label{eq:lamPCE}\\
\lambda_2\left(\ve{x}\right) &\approx \lambda^{\PC}_2\left(\ve{x};\ve{c}\right) 
= \exp \left(\sum_{\ve{\alpha}\in \ca_2} 
c_{2,\ve{\alpha}}\psi_{\ve{\alpha}}(\ve{x}) \right), \label{eq:lamPCE_lam2}
\end{align}
where $\ca_l$ ($l = 1,2,3,4$) is a finite set of selected basis 
functions for $\lambda_l$, and $c_{l,\ve{\alpha}}$'s are the coefficients. For 
the purpose of clarity, we explicitly express $\ve{c}$ in the PC approximations 
$\lambda^{\PC}_l\left(\ve{x};\ve{c}\right)$ to emphasize that $\ve{c}$ are the 
model parameters yet to be estimated from data. 

\subsection{GLaM constructions}
\label{sec:estimation}
We assume that our costly stochastic simulator is evaluated once for each point 
$\ve{x}^{(i)}$ of the experimental design $\cx$, and the associated model 
response $y^{(i)}$ is collected in $\cy$:
\begin{equation}
	\cx = \acc{\ve{x}^{(1)},\ldots,\ve{x}^{(N)}}, \quad \cy = 
	\acc{\cm_s\left(\ve{x}^{(1)},\omega^{(1)}\right),\ldots,
		\cm_s\left(\ve{x}^{(N)},\omega^{(N)}\right)}
\end{equation}
As already mentioned (and as emphasized by the notation $\omega^{(i)}$), no 
replications are required, and we do not control the random 
seed. To 
construct a GLaM from the available data $(\cx,\cy)$, both the truncated sets 
$\ve{\ca}$ of basis functions and the coefficients $\ve{c}$ shall be 
determined. In this section, we summarize the method proposed in 
\cite{ZhuSIAMUQ2021}, which is designed to achieve both purposes without the 
need for replications. 

Sometimes prior knowledge is available to set the basis functions. For example, 
when working with a standard linear regression problem, the data is supposed to 
be generated by
\begin{equation}
	Y = \beta_0 + \ve{X}\ve{\beta} + \epsilon,
\end{equation}
where $\epsilon$ has mean zero and is independent of $\ve{X}$. This case can be 
treated within the GLaM framework as follows: $\ca_1$ contains the constant and 
linear term, and $\ca_2$, $\ca_3$, $\ca_4$ contain only a constant term. Note 
that such a GLaM allows estimating the distribution of $\epsilon$, which is not 
required to be normal, whereas 
the usual linear regression framework assumes normally distributed 
$\epsilon$. 

However, in general there is no prior knowledge that would help select 
$\ve{\ca}$. Thus, we make the following assumptions to find appropriate 
hyperbolic truncation schemes defined in \Cref{eq:qnorm} for each $\lambda_i, 
i=1,\ldots,4$: 
\begin{enumerate}[label=(A\arabic*)]
	\item The response distribution of $Y(\ve{x})$ can be well-approximated by 
	a generalized lambda distribution;
	\item The shape of this distribution smoothly varies as a function of $\ve{x}$, so that the parameters $\lambda_i(\ve{x})$ can be well approximated by a low-order PCE
\end{enumerate}

Because the shape of a GLD is controlled by $\lambda_3$ and $\lambda_4$, the 
associated hyperbolic truncation schemes $\ca^{p,q,M}$ can be set with a small 
value of $p$, say $p=1$. 

Moreover, the parameters $\lambda_1(\ve{x})$ and $\lambda_2(\ve{x})$ mainly 
affect the variation of the mean $m(\ve{x})$ and of the variance $v(\ve{x})$ as 
a function of the input $\ve{x}$, respectively. As a result, 
they may require possibly larger degree $p$. To this end, we modify the 
feasible generalized least-squares (FGLS) \cite{Wooldridge2013} to find 
suitable truncation schemes for the mean and variance function modeled as
  
\begin{equation*}
m(\ve{x}) = \sum_{\ve{\alpha}\in \ca_{m}} 
c_{m,\ve{\alpha}}\psi_{\ve{\alpha}}(\ve{x}), \quad
v(\ve{x}) = \exp \left(\sum_{\ve{\alpha}\in \ca_v} 
c_{v,\ve{\alpha}}\psi_{\ve{\alpha}}(\ve{x}) \right).
\end{equation*}
 
Basically, FGLS iterates between a weighted least-square problem (WLS) to fit 
the mean function, and an ordinary least-square (OLS) analysis to estimate the 
variance function.

The details of the modified FGLS are presented in \Cref{alg:FGLS}. 
In this algorithm, the inputs $\ve{p}_1$ and $\ve{q}_1$ stand for
the set of candidate degrees and $q$-norms that are tested 
to expand $\lambda_1(\ve{x})$, 
respectively. The same notation apply to $\ve{p}_2$ and $\ve{q}_2$ for 
$\lambda_2(\ve{x})$. Indeed, because of the low cost of least-square analysis, 
various combinations of $p$ and $q$ are tested for both $\lambda_1(\ve{x})$ and 
$\lambda_2(\ve{x})$.

More precisely, AOLS denotes \emph{adaptive ordinary least-squares} with 
degree and $q$-norm adaptivity \cite{UQdoc_13_104,SudretJCP2011}. This 
algorithm first builds a series of PCEs, each of which is obtained by applying 
ordinary least-squares with a truncation scheme $\ca^{p,q,M}$ defined by a 
combination of $p\in \ve{p}$ and $q \in \ve{q}$. Then, it selects the PCE, 
therefore the associated truncation scheme, with the smallest 
\emph{leave-one-out} errors (see \cite{SudretJCP2011} for details). $\WLS$ 
denotes the use of weighted least-squares, which takes the estimated variance 
$\hat{\ve{v}}$ as weight to re-estimate $\ve{c}_{m}$. In this procedure, the 
truncation set $\ca_{m}$ for $m(\ve{x})$ is selected only once (before 
the loop), whereas a set of truncation schemes $\acc{\ca^i_{v}: i = 
1,\ldots,N_{\rm FGLS}}$ is obtained. 

We finally select the one with the smallest 
leave-one-out error as the final truncated set $\ca_{v}$ for $v(\ve{x})$. The 
number of iterations $N_{\rm FGLS}$ is defined by the user, typically $N_{\rm 
FGLS} = $ 5--10. After applying \Cref{alg:FGLS}, we set $\ca_1 = \ca_{m}$ and 
$\ca_2 = \ca_{v}$.

\begin{algorithm}[H]
	\caption{Modified feasible generalized least-squares}
	\label{alg:FGLS}
	\begin{algorithmic}[1]
		\STATE Input: $\left(\cx,\cy\right)$, $\ve{p}_1$, $\ve{q}_1$, 
		$\ve{p}_2$, $\ve{q}_2$
		\STATE Output: truncated sets for the mean and variance 
		function--$\ca_{m}$ and $\ca_{v}$
		\STATE $\ca_{m},\,\, \hat{\ve{c}}_{m} \gets 
		\AOLS\left(\cx,\cy,\ve{p}_1,\ve{q}_1\right)$
		\FOR{$i \gets 1,\ldots,N_{\rm FGLS}$}
		\STATE $\hat{\ve{m}} \gets \sum_{\ve{\alpha} \in 
			\ca_{m}}c_{m,\ve{\alpha}}\psi_{\ve{\alpha}}(\cx)$
		\STATE $\tilde{\ve{r}} \gets 2\log\left(\abs{\cy - 
			\hat{\ve{m}}}\right)$
		\STATE $\ca^i_{v},\,\,\hat{\ve{c}}_{v},\,\, \varepsilon^i_{\rm 
		LOO}\gets 
		\AOLS\left(\cx,\tilde{\ve{r}},\ve{p}_2,\ve{q}_2\right)$
		\STATE $\hat{\ve{v}} \gets \exp\left(\sum_{\ve{\alpha} \in 
			\ca_{v}}c_{v,\ve{\alpha}}\psi_{\ve{\alpha}}(\cx)\right)$
		\STATE $\hat{\ve{c}}_{m} \gets 
		\WLS\left(\cx,\cy,\ca_{m},\hat{\ve{v}}\right)$
		\ENDFOR
		\STATE $ i^* = \arg\min \acc{\varepsilon^i_{\rm LOO} : 
		i=1,\ldots,N_{\rm FGLS}}$ and $\ca_{v}\gets \ca^{i^*}_{v}$
	\end{algorithmic}
\end{algorithm}

Once the basis functions are selected, we use the maximum 
(conditional) likelihood estimator to estimate $\ve{c}$:
\begin{equation}\label{eq:joint}
\hat{\ve{c}} = \arg\min_{\ve{c} \in \cc} \mathsf{L}\left(\ve{c}\right),
\end{equation}
where $\mathsf{L}\left(\ve{c}\right)$ is the conditional negative log-likelihood
\begin{equation}\label{eq:nloglh}
\mathsf{L}\left(\ve{c}\right) = \sum_{i=1}^{N}-\log\left( 
f^{\GLD}\left(y^{(i)} ; 
\ve{\lambda}^{\PC}\left(\ve{x}^{(i)};\ve{c}\right)\right)\right),
\end{equation}
with $f^{\GLD}$ being the probability density function of the GLD defined in 
\Cref{eq:FKMLpdf}. 

The advantages of the proposed estimator are twofold. On the one hand, the 
simulator is required to be evaluated only once (but not limited to one) on 
each point of the experimental design. Thereby, replications are not 
necessary (yet possible), and the method is versatile in this respect. On the 
other hand, if the underlying computational model can be exactly represented by 
a GLaM for a specific choice of $\ve{c}$, the maximum likelihood estimator is 
\emph{consistent} (see proof in \cite{ZhuSIAMUQ2021}).

In practice, the evaluation of $\mathsf{L}(\ve{c})$ is not 
straightforward because the PDF of generalized lambda distributions does not 
have an explicit form: it is necessary to solve 
nonlinear equations as shown in \cref{eq:FKMLpdf}. Nevertheless, the nonlinear 
function $Q(u;\ve{\lambda})$ is monotonic and defined on $[0,1]$. Therefore, we 
proposed using the bisection method \cite{Burden2015} to efficiently solve the 
nonlinear equations.

\subsection{Sensitivity analysis with GLaMs}
\label{sec:SGLaM}
\subsubsection{Introduction}
\label{sec:SGLaM1}
The various Sobol' indices introduced in \Cref{eq:ClassSobol} and 
(\ref{eq:qSobol}) can be estimated by sampling from the conditional 
distribution $Y\mid \ve{X}_{\iu}$. Because of the specific format of the GLaM 
definition, such a sampling can be easily performed.

The generalized lambda distribution parameterizes the quantile function 
$Q(u;\ve{\lambda})$ (see \Cref{eq:FKML}), which can be seen as the inverse 
probability integral transform. In other words, the random variable 
$Q(U;\ve{\lambda})$ with $U\sim \cu(0,1)$ follows $\GLD(\ve{\lambda})$. As a 
result, sampling from a GLD is straightforward. We define the function 
$Q_{\GLaM}: (u,\ve{x}) \in [0,1]\times \cd_{\ve{X}} \mapsto \Rr$ by
\begin{equation}\label{eq:GLaMS}
	Q_{\GLaM}(u;\ve{x}) = Q(u;\ve{\lambda}^{\PC}(\ve{x})) = 
	\lambda^{\PC}_1(\ve{x}) + 
	\frac{1}{\lambda^{\PC}_2(\ve{x})}\left( 
	\frac{u^{\lambda^{\PC}_3(\ve{x})}-1}{\lambda^{\PC}_3(\ve{x})} - 
	\frac{(1-u)^{\lambda^{\PC}_4(\ve{x})}-1}{\lambda^{\PC}_4(\ve{x})}\right).
\end{equation}
$Q_{\GLaM}(U;\ve{x})$ is a so-called \emph{GLaM stochastic emulator} where $U 
\sim \cu(0,1)$ serves as a latent variable that introduces the internal source 
of randomness. Precisely, $Q_{\GLaM}(U;\ve{x})$ is a random variable following 
the surrogate response PDF for $\ve{X} = \ve{x}$, and $Q_{\GLaM}(u;\ve{x})$ 
provides its corresponding $u$-quantile. In other words, GLaM is a simple 
stochastic surrogate model with only one latent variable (namely, $U$); this surrogate behaves similarly 
to the original stochastic simulator in terms of the response distribution for 
any $\ve{x}$. 

\Cref{eq:GLaMS} emulates the conditional quantile function 
$Q_{Y\mid\ve{X}}(u;\ve{x})$ of the original model, so calculating Sobol' indices $S_{\iu}$ of the deterministic function $Q_{\GLaM}(u;\ve{x})$ can directly provide the classical Sobol' indices defined in \Cref{eq:ClassSobol}. Note that $Q_{\GLaM}$ also allows us to calculate classical Sobol' indices involving $U$, \eg $S_{U}$. However, since the surrogate approximates only the 
response distribution but cannot produce the trajectories, these Sobol' indices 
are not representative of those of the original model, \eg $S_{\ve{Z}}$ that
requires estimating $\Var{\Esp{Y\mid\ve{Z}}}$ (according to 
\Cref{eq:ClassSobol}), where the inner expectation $\Esp{Y\mid\ve{z}}$ is 
taken over a trajectory.

For QoI-based Sobol' indices in \Cref{eq:qSobol}, if the quantity of interest 
$q_{\GLaM}(\ve{x})$ can be directly calculated from generalized lambda 
distributions, we can just treat it as a classical surrogate 
model of $\QoI(\ve{x})$. This is the case for the mean $m(\ve{x})$ and the 
variance $v(\ve{x})$ (see \Cref{sec:pGLDs} for details). In addition, if 
$\QoI(\ve{x})$ is a $u$-quantile of the response distribution, \Cref{eq:GLaMS} 
is 
used directly. 

Finally, if it is impossible to evaluate analytically $q_{\GLaM}(\ve{x})$, we 
generate a large sample set from \Cref{eq:GLaMS} by sampling $U\sim 
\cu(0,1)$, and then use the sample statistic $\hat{q}_{\GLaM}(\ve{x})$ as a 
surrogate model for $\QoI(\ve{x})$.

\subsubsection{Monte Carlo estimates}
\label{sec:SGLaM2}

Because both $Q_{\GLaM}(u;\ve{x})$ and $q_{\GLaM}(\ve{x})$ are deterministic, 
we can use methods based on Monte Carlo simulations \cite{Homma1996} to 
estimate the considered Sobol' indices. Here, we illustrate the estimator 
suggested by Janon et al. \cite{Janon2013} for classical Sobol' indices 
estimations.

We first define two random variables $Y = Q_{\GLaM}(U;\ve{X}_{\iu},\ve{X}_{\sim 
\iu})$ and $Y_{\iu} = Q_{\GLaM}(\tilde{U};\ve{X}_{\iu},\tilde{\ve{X}}_{\sim 
\iu})$, where $\tilde{U}$ and $\tilde{\ve{X}}_{\sim \iu}$ are independent 
copies of $U$ and $\ve{X}_{\sim \iu}$. This indicates that $Y_{\iu}$ is 
correlated to $Y$ by using the same set of random variables $\ve{X}_{\iu}$ as 
argument. In addition, $Y$ and $Y_{\iu}$ follow the same distribution, and thus 
they share the same moments, \eg $\Esp{Y} = \Esp{Y_{\iu}}$, $\Esp{Y^2} = 
\Esp{Y^2_{\iu}}$.

Following Janon et al. \cite{Janon2013}, $S_{\iu}$ defined in 
\Cref{eq:ClassSobol} can be re-written as:
\begin{equation}\label{eq:covSobol}
	S_{\iu} = \frac{\Cov{Y,Y_{\iu}}}{\Var{Y}} = \frac{\Esp{Y\,Y_{\iu}} - 
	\left(\Esp{Y}\right)^2}{\Esp{Y^2} - \left(\Esp{Y}\right)^2} = 
	\frac{\Esp{Y\,Y_{\iu}} - 
	\left(\frac{1}{2}\Esp{Y+Y_{\iu}}\right)^2}{\frac{1}{2}\Esp{Y^2 + Y^2_{\iu}} 
	- 	\left(\frac{1}{2}\Esp{Y+Y_{\iu}}\right)^2}.
\end{equation}

We generate $N_{\MC}$ realizations of $Y$ and $Y_{\iu}$ by sampling 
(independently) $\ve{X}_{\iu}$, $\ve{X}_{\sim \iu}$, $U$, $\tilde{\ve{X}}_{\sim 
\iu}$, and $\tilde{U}$. The expectations in \Cref{eq:covSobol} 
can be estimated by sample statistics, which leads to
\begin{equation}\label{eq:pick-freeze}
\hat{S}_{\iu} = \frac{\frac{1}{N_{\MC}} \sum_{i=1}^{N_{\MC}} 
y^{(i)}\,y^{(i)}_{\iu} - 
\left(\frac{1}{2N_{\MC}} \sum_{i=1}^{N_{\MC}} 
\left(y^{(i)}+y^{(i)}_{\iu}\right)\right)^2}{\frac{1}{2N} 
\sum_{i=1}^{N_{\MC}} \left( \left(y^{(i)}\right)^2 + 
\left(y^{(i)}_{\iu}\right)^2 \right) - 
\left(\frac{1}{2N_{\MC}} \sum_{i=1}^{N_{\MC}} y^{(i)}+y^{(i)}_{\iu}\right)^2},
\end{equation}
where $y^{(i)} = 
Q_{\GLaM}\left(u^{(i)};\ve{x}_{\iu},\ve{x}_{\sim \iu}\right)$ 
and $y_{\iu}^{(i)} = 
Q_{\GLaM}\left(\tilde{u}^{(i)};\ve{x}_{\iu},\tilde{\ve{x}}_{\sim 
	\iu}\right)$ are the $i$-th realizations of $Y$ and $Y_{\iu}$, respectively.

For QoI-based Sobol' indices defined in \Cref{eq:qSobol}, we 
follow the same procedure by replacing $Q_{\GLaM}$ by $q_{\GLaM}$.

\subsubsection{PCE-based estimates}
\label{sec:SGLaM3}
As discussed in \Cref{sec:SGLaM1}, estimating the considered two types of 
Sobol' indices of a GLaM surrogate model is reduced to studying two 
deterministic functions $Q_{\GLaM}(u;\ve{x})$ and $q_{\GLaM}(\ve{x})$. 
According to the definition in \Cref{sec:Sobol}, $\ve{X}$ has mutually 
independent components, 
which are also independent of $U \sim \cu(0,1)$. Both functions can be 
represented by polynomial chaos expansions (see \Cref{sec:PCE}): 
\begin{equation}
	\begin{split}
		Q_{\GLaM}(u;\ve{x}) &\approx Q^{\PC}_{\GLaM}(u;\ve{x}) = 
		\sum_{\ve{\alpha}\in 
			\ca^Q}c^Q_{\ve{\alpha}}\psi^Q_{\ve{\alpha}}(u,\ve{x}), \qquad 
			\text{ and }\\
		q_{\GLaM}(\ve{x}) &\approx q^{\PC}_{\GLaM}(\ve{x}) = 
		\sum_{\ve{\alpha}\in 
			\ca^q}c^q_{\ve{\alpha}}\psi_{\ve{\alpha}}(\ve{x}),
	\end{split}
\end{equation}
where $\ca^Q \subset \Nn^{M+1}$ and $\ca^q \subset \Nn^{M}$ are the truncated 
sets defining the basis functions $\psi_{\ve{\alpha}}(\ve{x})$'s and 
$\psi^Q_{\ve{\alpha}}(u,\ve{x})$'s, respectively, as discussed in 
\Cref{eq:PCE_basis}. Note that each multi-index in $\ca^Q$ has a dimension 
$M+1$ because of the additional variable $u$, and the univariate basis 
functions of $u$ in $\psi_{\ve{\alpha}}^Q(u,\ve{x})$ are Legendre polynomials 
\cite{Xiu2002}. The advantage of using a PCE surrogate is that its Sobol' 
indices (of any order) can be analytically calculated by post-processing its 
coefficients \cite{SudretRESS2008b}. 

Several methods have been developed to construct PCEs for deterministic 
functions with given basis functions, such as the projection method 
\cite{Ghanembook2003} and ordinary least-squares \cite{Berveiller2006}. To both
determine the truncated set and estimate the associated coefficients, we opt 
for the hybrid-LAR algorithm \cite{SudretJCP2011}. This method selects 
the most 
important basis functions among a candidate set, before ordinary least-squares 
is used to compute the coefficients. The selection procedure of the algorithm 
is based on \emph{least angle regression} (LAR) \cite{Efron2004}. 

Practically, we first generate $N_{\PC}$ samples by sampling $\ve{X}$ and $U$. 
They are used to evaluate the target function $Q_{\GLaM}(u;\ve{x})$ or 
$q_{\GLaM}(\ve{x})$ to obtain the associated model responses. Then, we apply 
the hybrid-LAR algorithm with the generated data to construct the PCE 
surrogate. Finally, the Sobol' indices are calculated by post-processing the PC 
coefficients.

In the following examples, we use the PCE-based estimates for the Sobol' 
indices of the GLaM surrogate model, instead of performing Monte Carlo 
simulations, as the accuracy of the former turned out to be extremely good.

\section{Examples}
\label{sec:examples}

In this section, we illustrate the performance of GLaMs for global sensitivity 
analysis on an analytical example and two case studies. We focus on the 
classical first-order Sobol' indices and QoI-based 
total Sobol' indices. The choice of the QoI depends on the 
focus of the example. To characterize the examples, we define the 
\emph{signal-to-noise ratio} of a stochastic simulator by
\begin{equation}
\SNR = \frac{\Var{\Esp{Y\mid\ve{X}}}}{\Esp{\Var{Y\mid\ve{X}}}} = 
\frac{\Var{m(\ve{X})}}{\Var{Y}-\Var{m(\ve{X})}}.
\end{equation}
This quantity gives the ratio between the variance of $Y$ 
explained by the mean function $m(\ve{x})$ and the remaining variance.

We use Latin hypercube sampling \cite{McKay1979} to generate the 
experimental design. The stochastic simulator is evaluated only once on 
each combination of input parameters. The associated output values are used to 
construct surrogates with the proposed estimation procedure introduced in 
\Cref{sec:estimation}. 

To assess the overall surrogate quality, we define the error measure
\begin{equation}\label{eq:errorrd}
\varepsilon_{Q} \eqdef \frac{\Esp{\left(Q_{Y\mid\ve{X}}(U;\ve{X}) - 
		Q_{\GLaM}(U;\ve{X})\right)^2}}{\Var{Q_{Y\mid\ve{X}}(U;\ve{X})}} = 
\frac{\Esp{\left(Q_{Y\mid\ve{X}}(U;\ve{X}) - 
		Q_{\GLaM}(U;\ve{X})\right)^2}}{\Var{Y}}
\end{equation}
where $Q_{Y\mid\ve{X}}(u;\ve{x})$ is the conditional quantile function of the 
model, $Q_{\GLaM}(u;\ve{x})$ is that of the GLaM following the definition in 
\Cref{eq:GLaMS}, and $U \sim \cu(0,1)$. This error has a form similar 
to the 
Wasserstein distance between probability measures \cite{Villani2008}. In 
addition, we also define an error measure to assess the accuracy of estimating 
the quantity of interest $\QoI(\ve{x})$ whose approximation by GLaM is denoted 
by $q_{\GLaM}(\ve{x})$
\begin{equation}\label{eq:errorq}
\varepsilon_{q} \eqdef \frac{\Esp{\left(\QoI(\ve{X}) - 
		q_{\GLaM}(\ve{X})\right)^2}}{\Var{\QoI(\ve{X})}}.
\end{equation}
The expectations in \Cref{eq:errorrd} and \Cref{eq:errorq} are estimated by 
averaging the error over a test set $\cx_{\rm test}$ of size $10^5$.

Experimental designs of various size $N$ are investigated to study the 
convergence of the proposed method. For each size, 50 independent realizations of these experimental designs are carried out to account for statistical uncertainty in the random design. As a consequence, estimates for each scenario are represented by box plots. 

\subsection{A three-dimensional toy example}
\label{sec:ex1}
The first example is defined as follows:
\begin{equation}\label{ex:logn}
Y(\ve{x},\omega)= \sin(x_1) + 7\sin^2\left(x_2\right) + 
\exp\left(\frac{x_1}{\pi} + x_3\,Z(\omega)\right)
\end{equation}
where $X_1, X_2 \sim \cu(0,2\pi)$, $X_3\sim\cu(0.25,0.75)$ are independent 
input variables, and $Z \sim \cn(0,1)$ denotes the latent variable that 
introduces the intrinsic randomness. The response 
distribution is a shifted lognormal distribution: the shift is equal to 
$\sin(x_1)+7\sin^2\left(x_2\right)$ and the lognormal distribution 
is parameterized by $\cl\cn\left(\frac{x_1}{\pi},x_3\right)$. As a 
result, this stochastic simulator has a nonlinear location function and a 
strong heteroskedastic effect: the variance varies between 0.069 and 72.35. 
Besides, this example has a mild signal-to-noise ratio $\SNR = 1.4$. This 
implies that the input variables can explain around 58\% of the total variance 
of $Y$ (\ie $S_{\acc{1,2,3}} = 0.58$).
\par
\begin{figure}[!htbp]
	\centering
	\begin{subfigure}[!b]{.45\linewidth}
		\centering
		\includegraphics[height=0.75\linewidth, keepaspectratio]{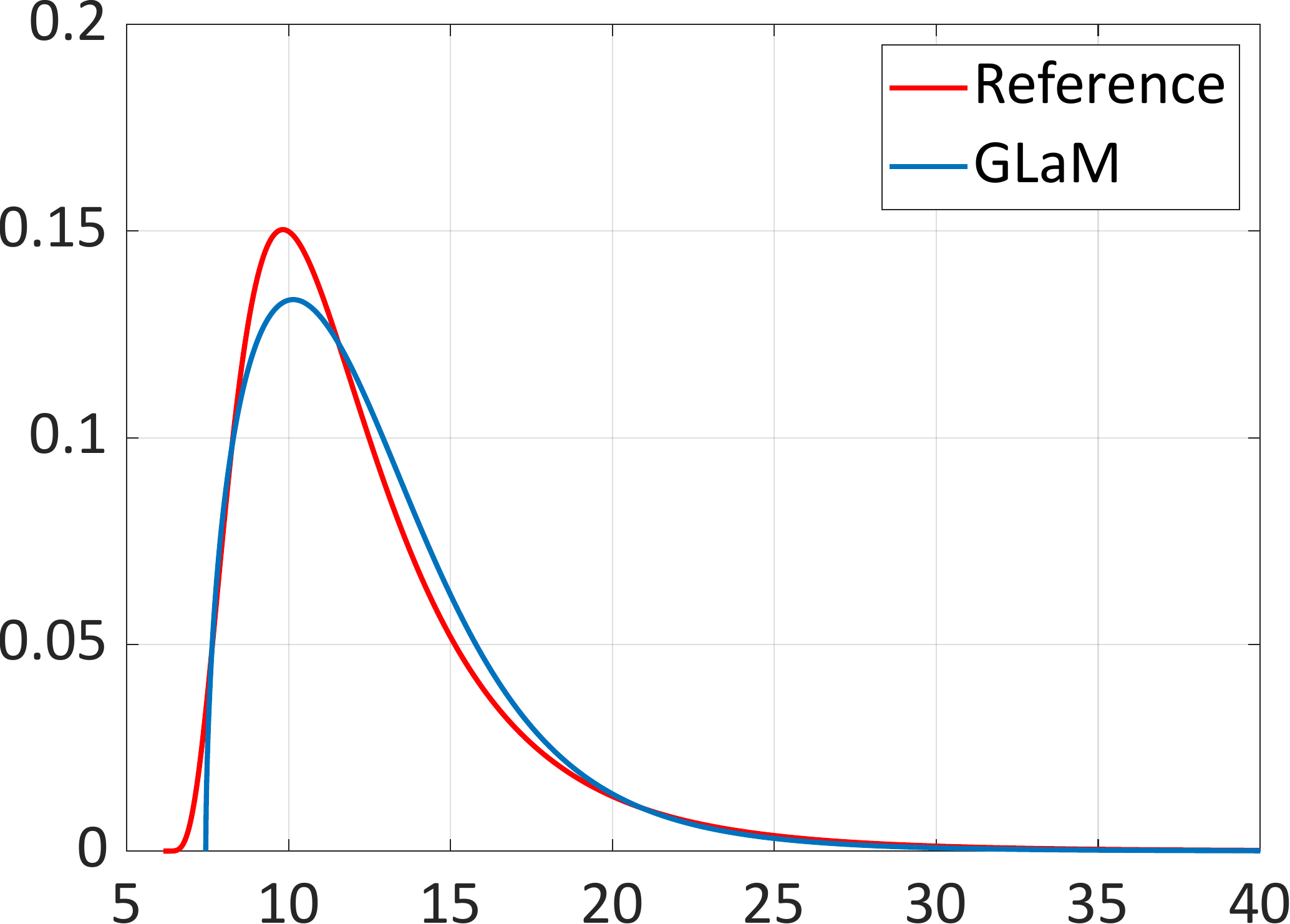}
		\caption{PDF for $\ve{x} = (5\pi/3,\pi/2,0.6)$}
		\label{subfig:lognpdf1}
	\end{subfigure}
	\hspace{0.5cm}
	\begin{subfigure}[!b]{.45\linewidth}
		\centering
		\includegraphics[height=0.75\linewidth, keepaspectratio]{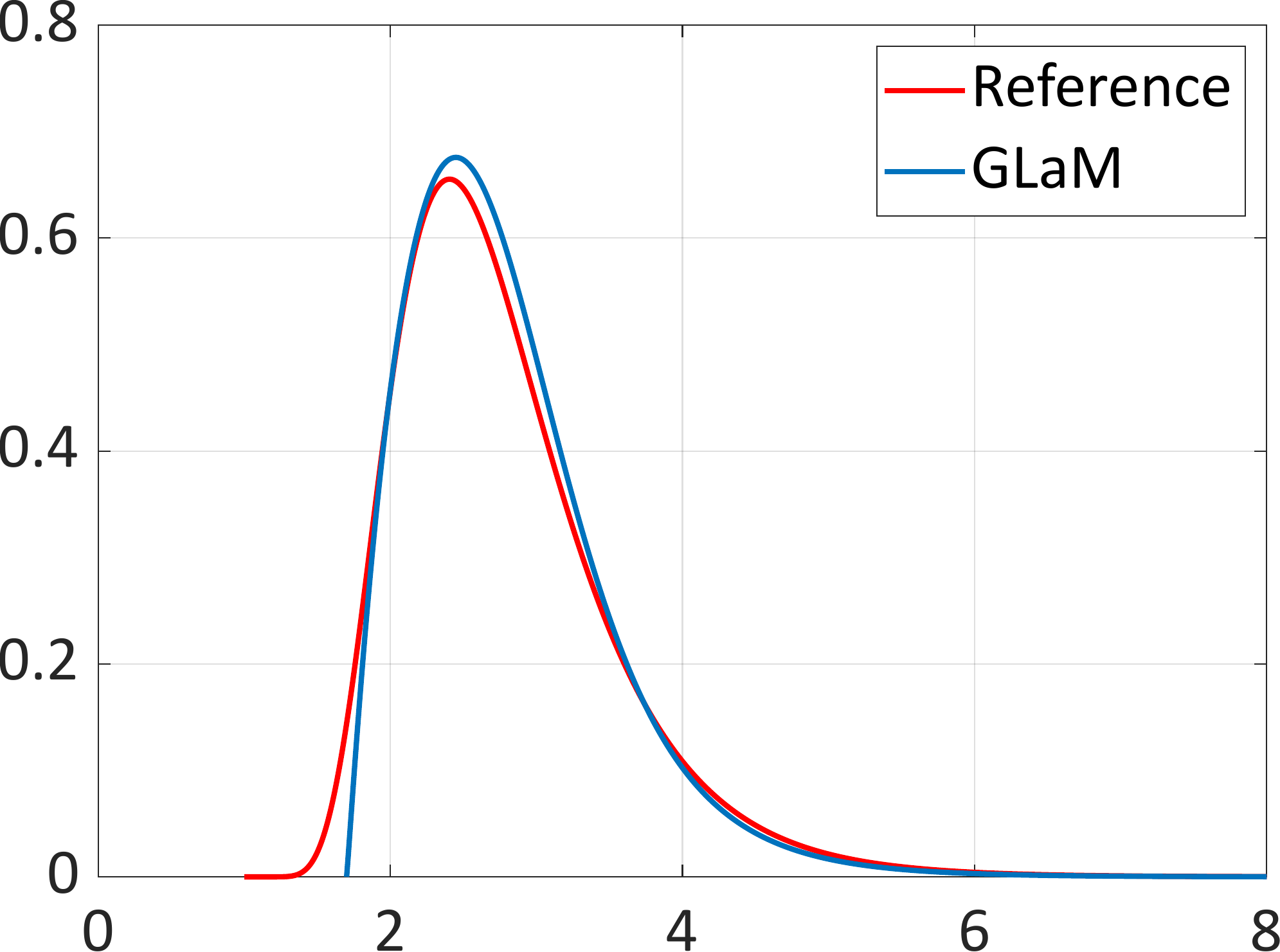}
		\caption{PDF for $\ve{x} = (\pi/3,\pi,0.4)$}
		\label{subfig:lognpdf2}
	\end{subfigure}
	\caption{Toy example -- Emulated response PDFs, $N=1{,}000$}.
	\label{fig:lognpdf}
\end{figure}
\par
\Cref{fig:lognpdf} compares the PDFs predicted by a GLaM built on an 
experimental design of $N=1{,}000$ with the reference response PDFs of the 
simulator. The results show that the developed algorithm correctly identifies 
the shape of the underlying shifted lognormal distribution. Moreover, the 
PDF supports and tails are also accurately approximated.
\par
We consider the differential entropy $h(\ve{x})$ \cite{AzziSens2020} as the QoI 
in this example. Because the analytical response distribution and entropy are 
known, we investigate the convergence of GLaM in terms of the conditional 
quantile function estimation \Cref{eq:errorq} and the entropy estimation 
\Cref{eq:errorrd}. The size of experimental design varies among $N \in 
\acc{250 ; 500 ; 1{,}000 ; 2{,}000 ; 4{,}000}$. Note that the entropy 
of a GLD does not have a closed form. Therefore, we use $10^4$ Monte Carlo 
samples to estimate this quantity of a GLaM for each $\ve{x}$ in the test set. 

In addition, we consider another model where we approximate the 
response distribution with a normal distribution. The mean and variance (as 
functions of $\ve{x}$) for such an approximation are chosen as the true mean 
and variance of the original. In other words, this model represents the ``oracle'' of
Gaussian-type mean-variance models.
\par
\begin{figure}[!htbp]
	\centering
	\begin{subfigure}[!b]{.45\linewidth}
		\centering
		\includegraphics[height=1\linewidth, 
		keepaspectratio]{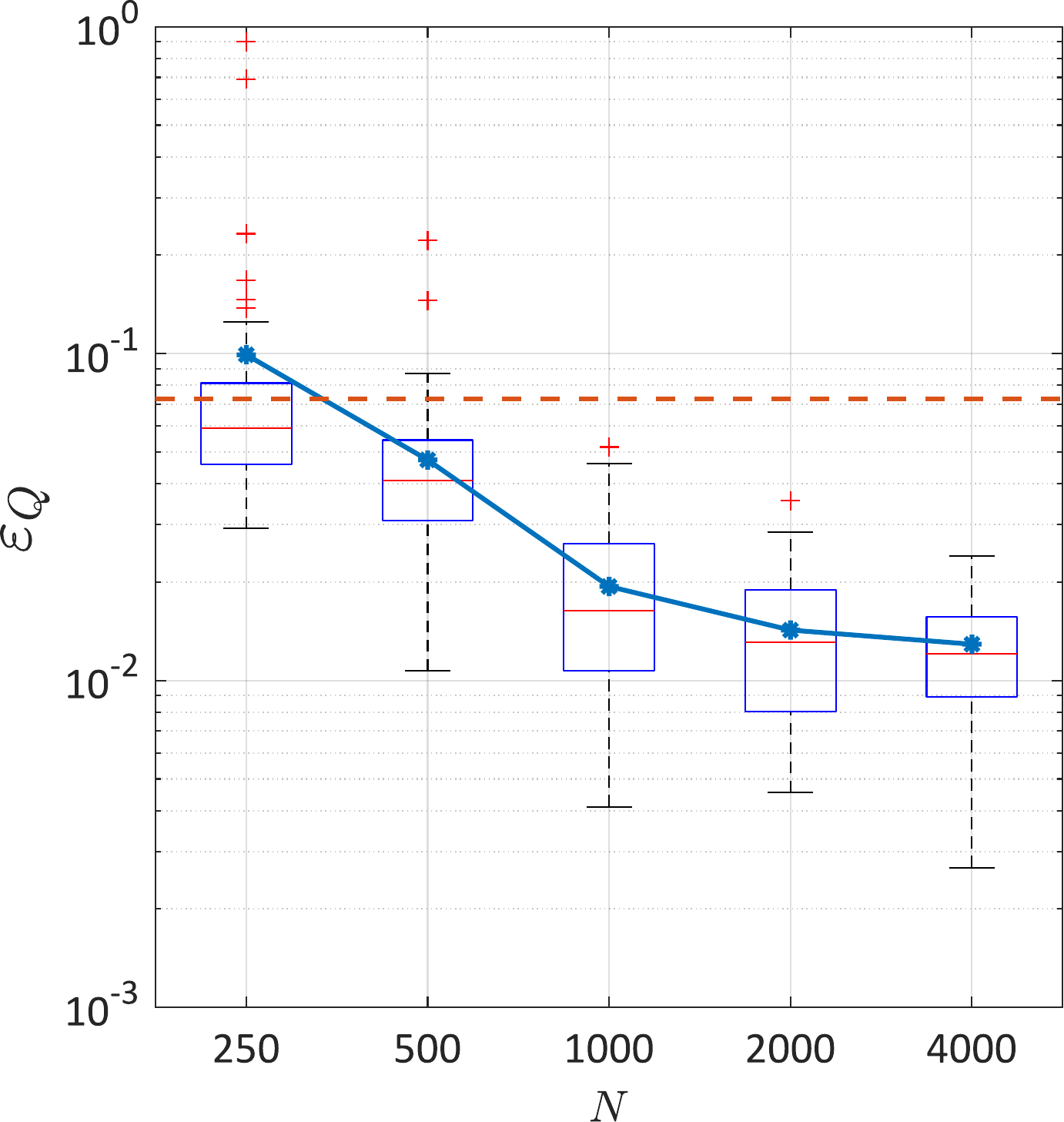}
		\caption{Conditional quantile function estimation}
		\label{fig:lognconv1}
	\end{subfigure}
\hspace{0.5cm}
	\begin{subfigure}[!b]{.45\linewidth}
	\centering
	\includegraphics[height=1\linewidth, 
	keepaspectratio]{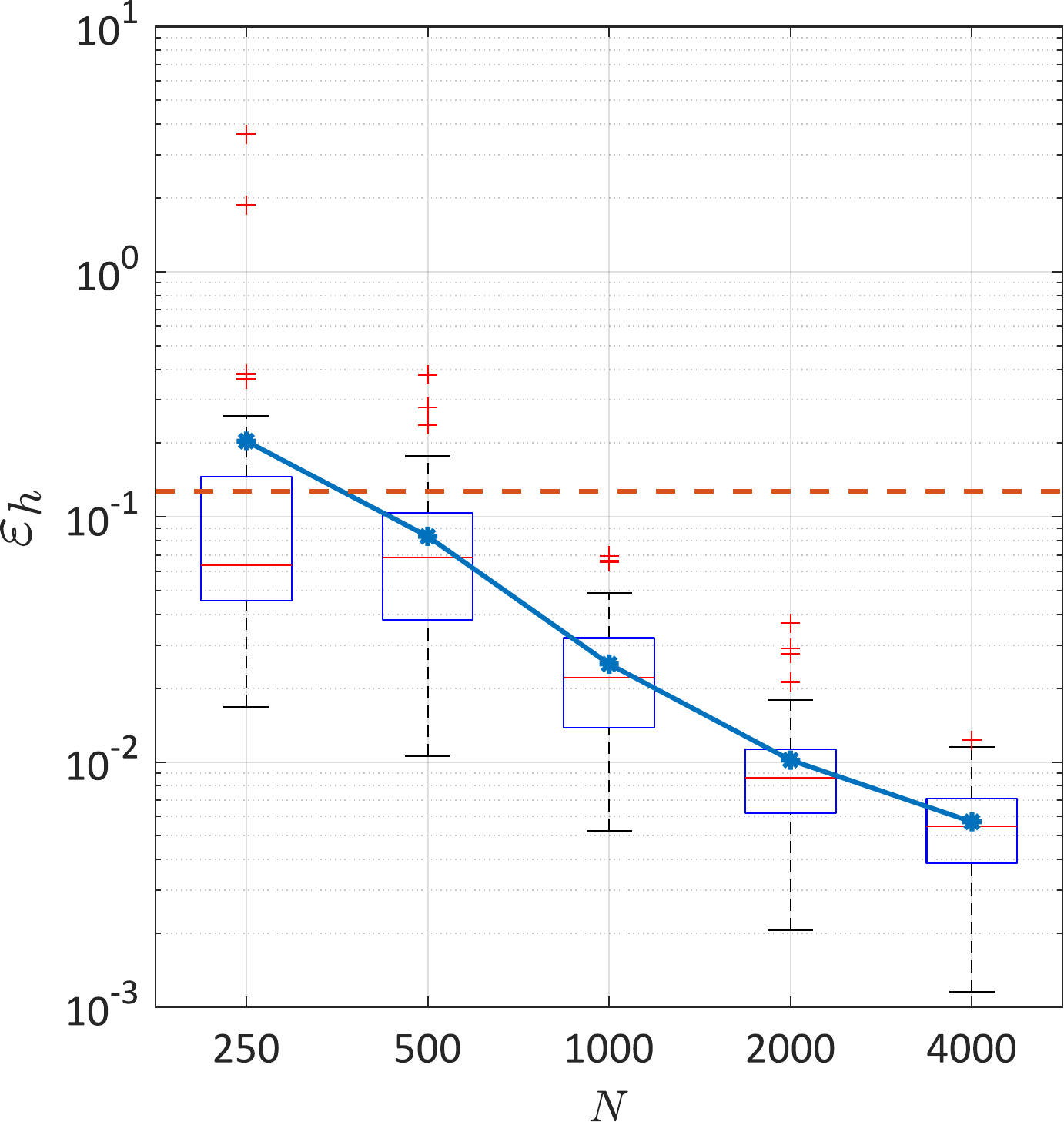}
	\caption{Entropy estimation}
	\end{subfigure}	
	\caption{Toy example -- Convergence study. The blue lines denote the 
	errors averaged over 50 repetitions of the full analysis. The red 
	dash lines are the corresponding errors of the model assuming that the 
	response distribution is normal with the true mean and variance}
	\label{fig:lognconv}
\end{figure}

The results are summarized in \Cref{fig:lognconv}. The 
proposed method exhibits a clear convergence with respect to $N$ for both 
$Q(u;\ve{x})$ and $h(\ve{x})$ estimations. We observe in \Cref{fig:lognconv1} 
that the decay of $\varepsilon_{Q}$ has two regimes separated by 
$N=1{,}000$. For a small $N$, the error coming from 
the use of finite samples dominates the estimate accuracy. When we consider a 
large data set, the error mainly comes from the model mispecification, because 
the stochastic simulator cannot be exactly represented by a GLaM. This 
phenomenon is not significant for the entropy estimation, which demonstrates a 
relatively consistent decay. 

The  accuracy of the oracle normal 
approximation is reported with red dash lines in \Cref{fig:lognconv}. The 
error shown is only due to model misspecifications (because the true response distribution is not Gaussian) since we use the underlying true 
mean and variance. For both measures, 
the medians of the errors of GLaMs built on $N=250$ model runs are smaller than 
those of the normal approximation. For $N\geq 1{,}000$, the GLaM clearly 
outperforms the oracle of Gaussian-type mean-variance models. This example 
illustrates the limits of such Gaussian-type models in practice.

Finally, the errors of GLaMs are below $0.05$ 
for $N\geq 1{,}000$ indicating that the 
surrogate is able to explain over 95\% of the variance of the target functions.

\begin{figure}[!htbp]
	\centering
	\includegraphics[width=0.95\linewidth, 
	keepaspectratio]{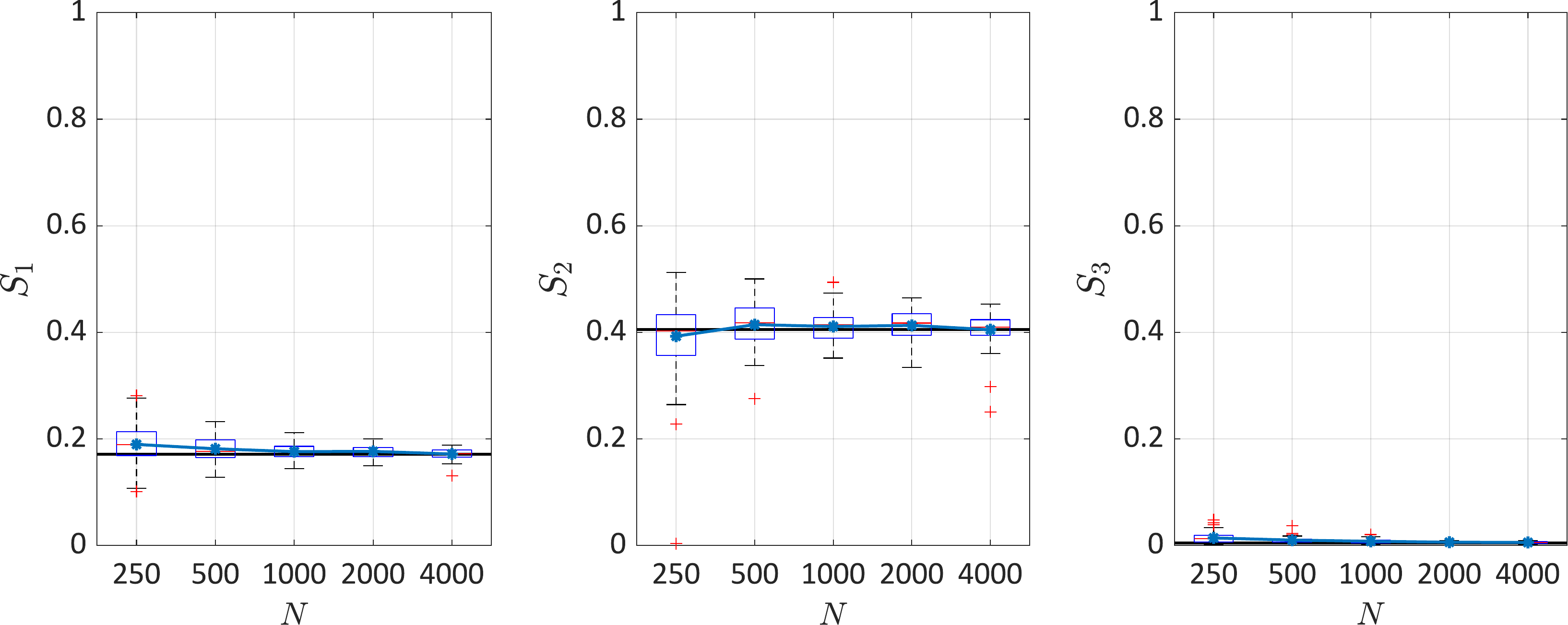}
	\caption{Estimation of the classical first-order Sobol' indices. The black 
		lines are the reference values, and the blue lines denote the average 
		values of 50 repetitions of the full analysis.}
	\label{fig:lognsen1}
\end{figure}

\begin{figure}[!htbp]
	\centering	
	\includegraphics[width=0.95\linewidth, 
	keepaspectratio]{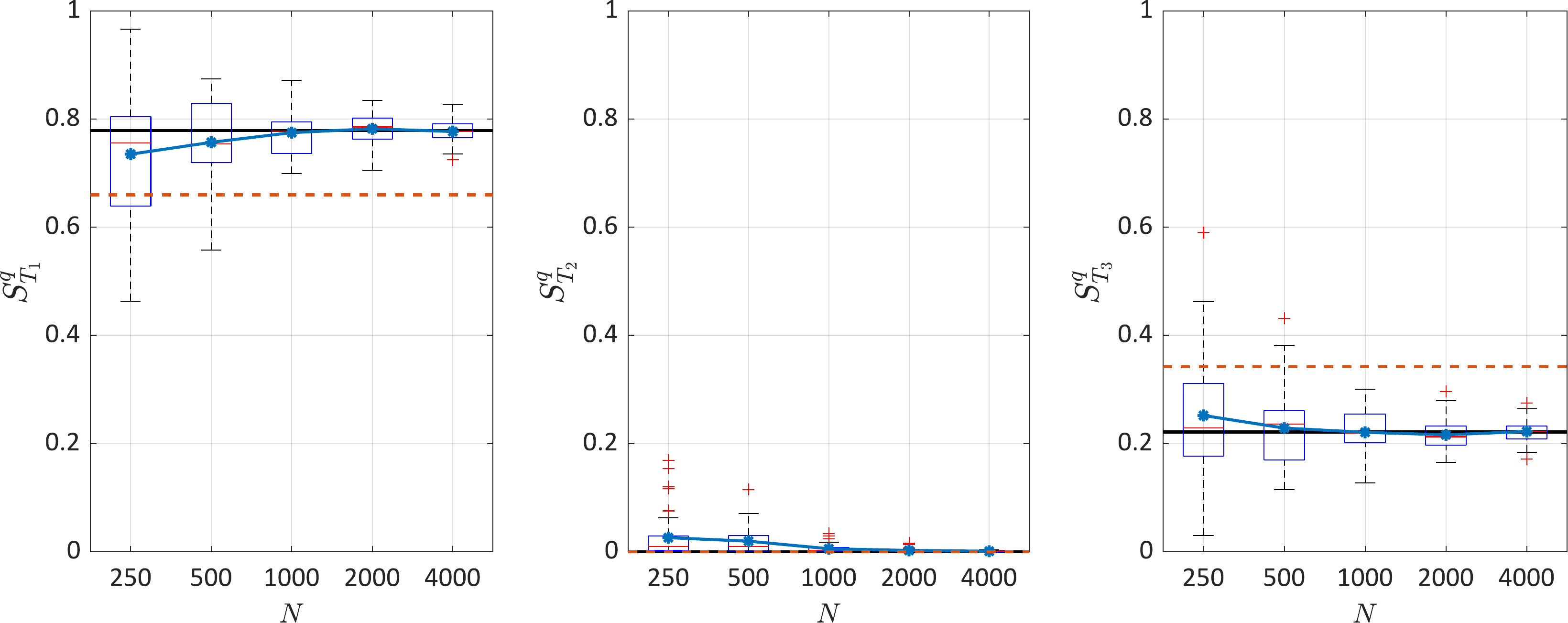}
	\caption{Estimation of the entropy-based total Sobol' indices. The black 
		lines are the reference values, and the blue lines denote the average 
		values of 50 repetitions of the full analysis. The red 
		dash lines correspond to the indices calculated from the normal 
		approximation using the true mean and variance}
	\label{fig:lognsen2}
\end{figure}

For sensitivity analysis, we focus on the classical first-order and the 
entropy-based total Sobol' indices. \Cref{fig:lognsen1} and 
\Cref{fig:lognsen2} show the convergence of GLaMs for estimating these 
quantities of each input variable. The reference values are derived from 
\Cref{ex:logn}. As shown by the two figures, this toy example is designed to 
have $X_2$ as the most important variable according to the classical 
first-order Sobol' indices, which also indicates that $X_2$ contributes the 
most to the variance of the mean function $m(\ve{X})$. In contrast, it has zero 
effect to the entropy. In comparison, $X_1$ is the dominant variable for the 
variation of the entropy $h(\ve{X})$. Because $X_3$ mainly controls the shape 
of the response distribution (especially the right tail), it has a minor 
first-order effect to the mean function, which leads to a very small value of 
$S_3$. In contrast, the entropy depends on the distribution shape, and thus 
$S^h_{T_3}$ is not negligible. The results reveal that GLaMs capture this 
characteristic and yield accurate estimates for both classical Sobol' indices 
and entropy-based Sobol' indices. 

Similar to \Cref{fig:lognconv}, we 
also reported the sensitivity indices calculated by using Gaussian 
approximations with the true mean and variance. Because the classical 
first-order indices depend only on the mean and variance functions, the oracle 
Gaussian model gives the exact values. Therefore, we showed only
the results for the entropy-based Sobol' indices in \Cref{fig:lognsen2}. It is 
clear that the Gaussian approximation with the true mean and variance 
demonstrates a significant bias. In contrast, GLaMs show nearly no bias and 
approximate much more accurately the reference values.

\subsection{Heston model}
\label{sec:Heston}
In this example, we perform the global sensitivity analysis for a Heston 
model used in mathematical finance \cite{Heston1993}. The Heston model 
describes the evolution of a stock price $Y_t$. It is an extension of the 
geometric Brownian motion by modeling the volatility as a stochastic process 
$v_t$, instead of considering it as constant. Hence, the Heston model is a 
stochastic volatility model and consists of two coupled stochastic differential 
equations:
\begin{equation}\label{eq:Heston}
\begin{split}
\D Y_t &= \mu Y_t \D t + \sqrt{v_t}Y_t \D W^1_t , \\
\D v_t &= \kappa(\theta - v_t) \D t + \sigma \sqrt{v_t} \D W^2_t ,
\end{split}
\end{equation}
with
\begin{equation}
	\Esp{\D W^1_t \, \D W^2_t} = \rho \D t,
\end{equation}
where $W^1_t$ and $W^2_t$ are two Wiener processes with correlation coefficient 
$\rho$, which introduce the intrinsic randomness of the 
stochastic model. The model parameters 
$\ve{x}=(\mu,\kappa,\theta,\sigma,\rho,v_0)$ are summarized in 
\Cref{tab:Hestonv}. The range of the last five input variables are selected 
based on the parameters calibrated from real data (S\&P 500 and Eurostoxx 50) 
\cite{Rouah2013}. The range of the first variable $\mu$ is set to $[0,0.1]$ to 
take the uncertainty of the expected return rate into account. Without loss of 
generality, we set $Y_0 = 1$.

\begin{table}[!htbp]
	\centering
	\caption{Parameters of the Heston model}
	\label{tab:Hestonv}
	\begin{tabular}{l l l}
		\hline
		Variable & Description & Distribution \\
		$\mu$ & Expected return rate & $\cu(0,0.1)$\\
		$\kappa$ & Mean reversion speed of the volatility & $\cu(0.3,2)$\\
		$\theta$ & Long term mean of the volatility & $\cu(0.02,0.07)$\\
		$\sigma$ & Volatility of the volatility & $\cu(0.2,0.4)$\\
		$\rho$ & Correlation coefficient between $\D W^1_t$ and $\D 
		W^2_t$ & $\cu(-1,-0.5)$\\
		$v_0$ & Volatility at time 0 & $\cu(0.02,0.07)$\\
		\hline
	\end{tabular}
\end{table}

In this example, we are interested in the stock price after one year, \ie 
$Y_t(\ve{x})$ with $t=1$. A closed form 
solution to \Cref{eq:Heston} is 
generally not available. To get samples of $Y_1(\ve{x})$, we simulate the 
entire time evolution of $Y_t$ and $v_t$ for a given $\ve{x}$ using Euler 
integration scheme with $\Delta t = 0.001$ over the time interval $[0,1]$. Note 
that when simulating the 
bivariate process $(Y_t, v_t)$, a problem may happen: since $v_t$ follows a 
Cox–Ingersoll–Ross process \cite{Rouah2013}, the simulation scheme can 
generate negative values for $v_t$. To overcome the problem, we apply the full 
truncation scheme, which replaces the update of $v_t$ by 
$\max\left(v_t,0\right)$ \cite{Rouah2013}. 

\begin{figure}[!htbp]
	\centering
	\begin{subfigure}[!b]{.49\linewidth}
		\centering
		\includegraphics[height=0.73\linewidth, 
		keepaspectratio]{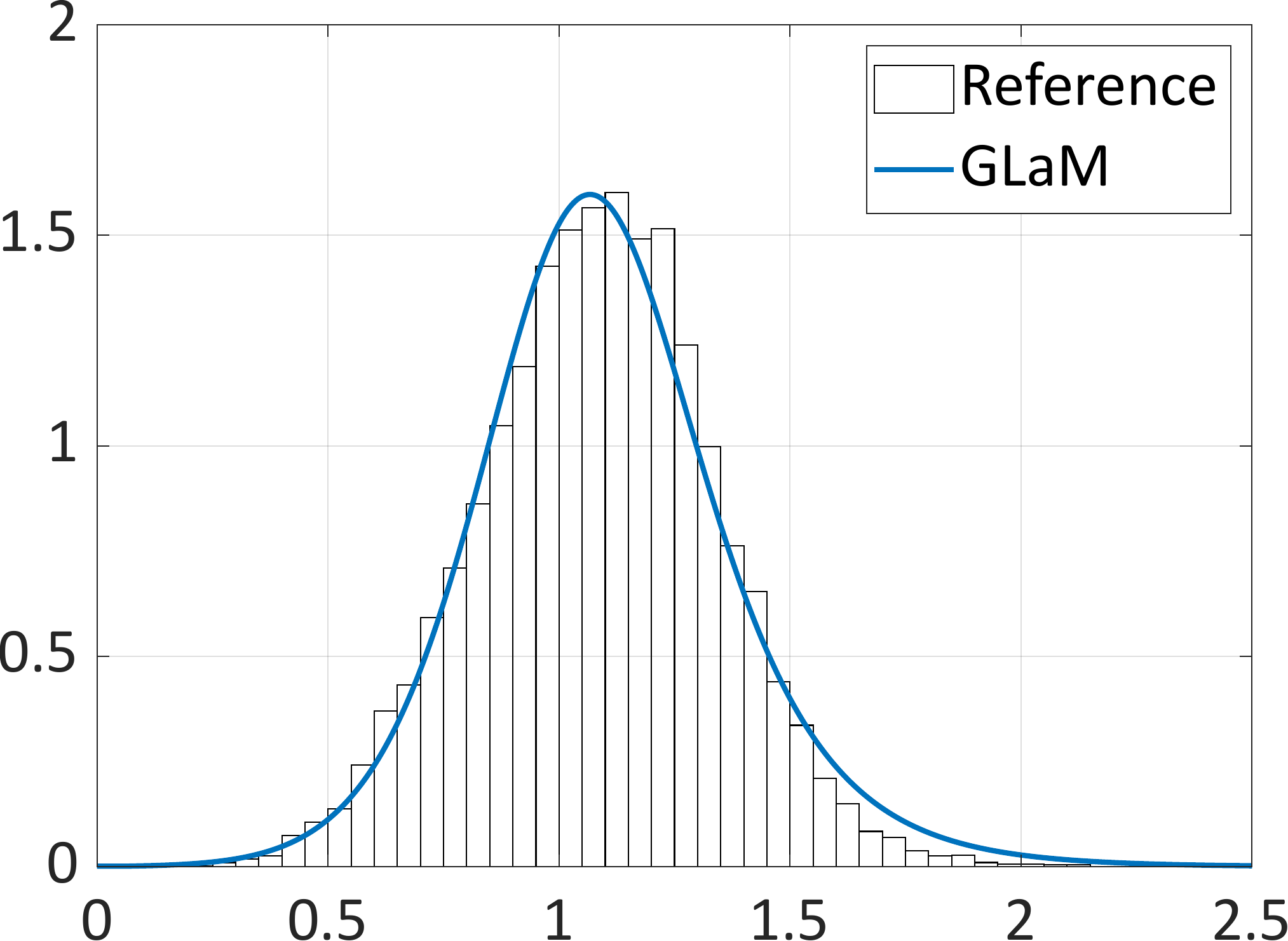}
		\caption{PDF for $\ve{x} = (0.08,1.6,0.06,0.35,-0.6,0.06)$}
		\label{subfig:Hestonpdf1}
	\end{subfigure}
%	\hspace{0.05cm}
	\begin{subfigure}[!b]{.49\linewidth}
		\centering
		\includegraphics[height=0.73\linewidth, 
		keepaspectratio]{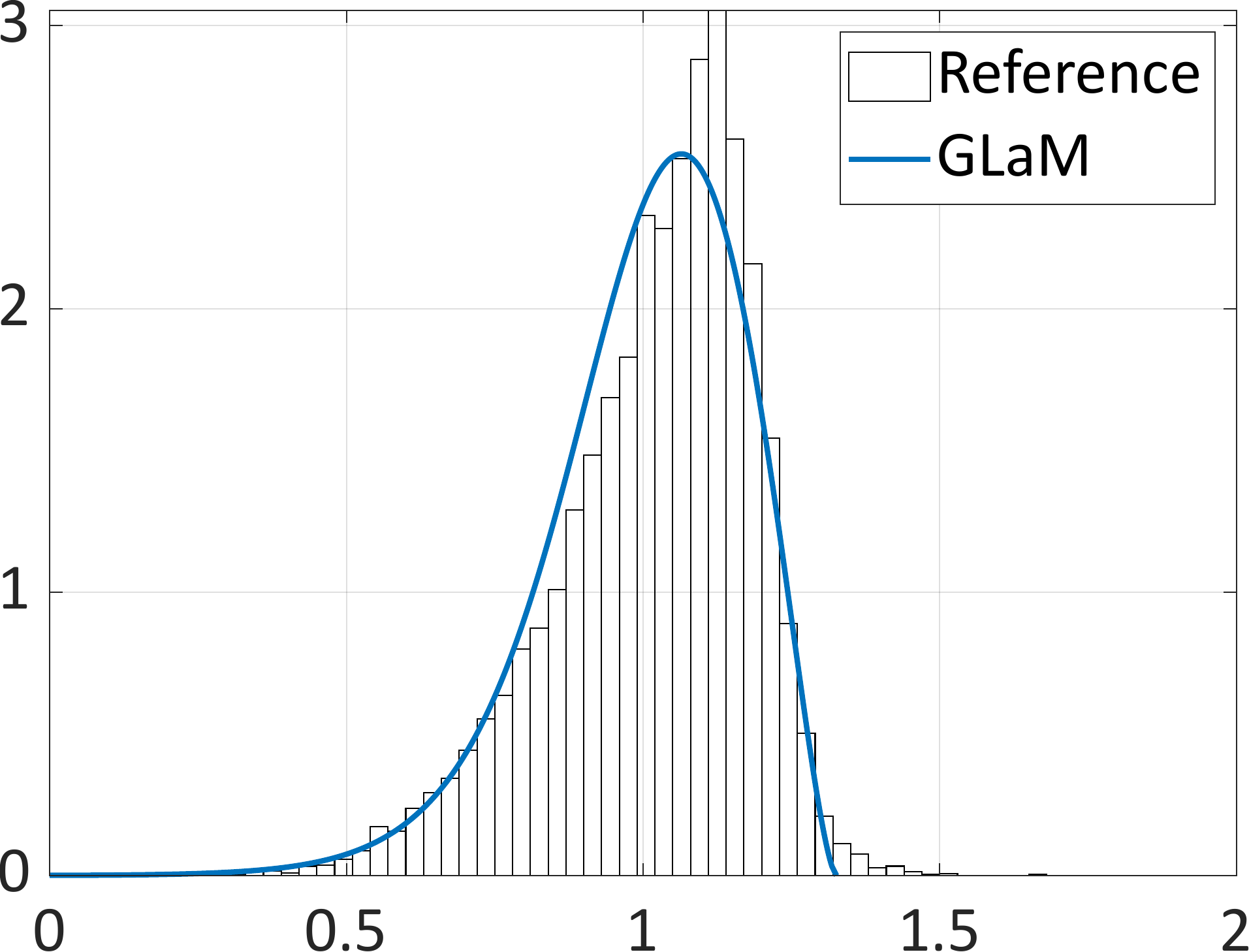}
		\caption{PDF for $\ve{x} = (0.02,0.5,0.03,0.25,-0.8,0.03)$}
		\label{subfig:Hestonpdf2}
	\end{subfigure}
	\caption{Heston model -- Emulated response PDFs, $N=2{,}000$}
	\label{fig:Hestonpdf}
\end{figure}

\Cref{fig:Hestonpdf} shows two response PDFs predicted by a surrogate built 
upon $N=2{,}000$ model runs. The reference 
histograms are obtained from 
$10^4$ repeated model runs with the same input parameters. We observe that 
the variance of the response distribution is not constant, \eg 0.065 and 0.027 
for the two illustrated PDFs. Moreover, the PDF shape varies: it 
changes from symmetric to left-skewed distributions depending on the model 
parameters. This would be difficult to approximate with a simple distribution 
family such as normal or lognormal. In contrast, GLaMs are able to accurately 
capture this shape variation, because of the flexibility of generalized lambda 
distributions.

Even though a closed form distribution of $Y_1(\ve{x})$ does not exist, the 
mean function $m(\ve{x}) = \Esp{Y_1(\ve{x})}$ can be derived analytically:
\begin{equation}\label{eq:HestonMean}
m(\ve{x}) = \exp(x_1) = \exp(\mu).
\end{equation}
As a result, we use $\varepsilon_{m}$ defined in \Cref{eq:errorq} with 
$\QoI(\ve{x}) = m(\ve{x})$ to assess the convergence of the surrogate. In 
addition, we also consider the expected payoff of an European call 
option. The payoff $C(\ve{x})$ and the expected payoff $m_C(\ve{x})$ of an 
European call option are defined by
\begin{equation}\label{eq:payoff}
\begin{split}
	C(\ve{x}) &= \max\acc{0,Y_1(\ve{x})-K}, \\
	m_C(\ve{x}) &= \Esp{C(\ve{x})},
\end{split}
\end{equation}
where $K$ is the \emph{strike price} and set to $1$ in the following analysis. 
In finance, $m_C$ not only is important for making investment decisions but 
also has a very similar form to the option price \cite{Shreve2004}. For the 
Heston model, numerical methods based on the Fourier transform have been 
developed to calculate the expected payoff without the need for Monte Carlo 
simulations \cite{Heston1993}. For the GLaM surrogate, this quantity can also 
be calculated numerically (see \Cref{sec:pGLDs}). As a second performance 
index, we compute the associated error denoted by $\varepsilon_C$ 
(\Cref{eq:errorq}) for the convergence study.
\par
\begin{figure}[!htbp]
	\centering
	\begin{subfigure}[!b]{.45\linewidth}
		\centering
		\includegraphics[height=1\linewidth, 
		keepaspectratio]{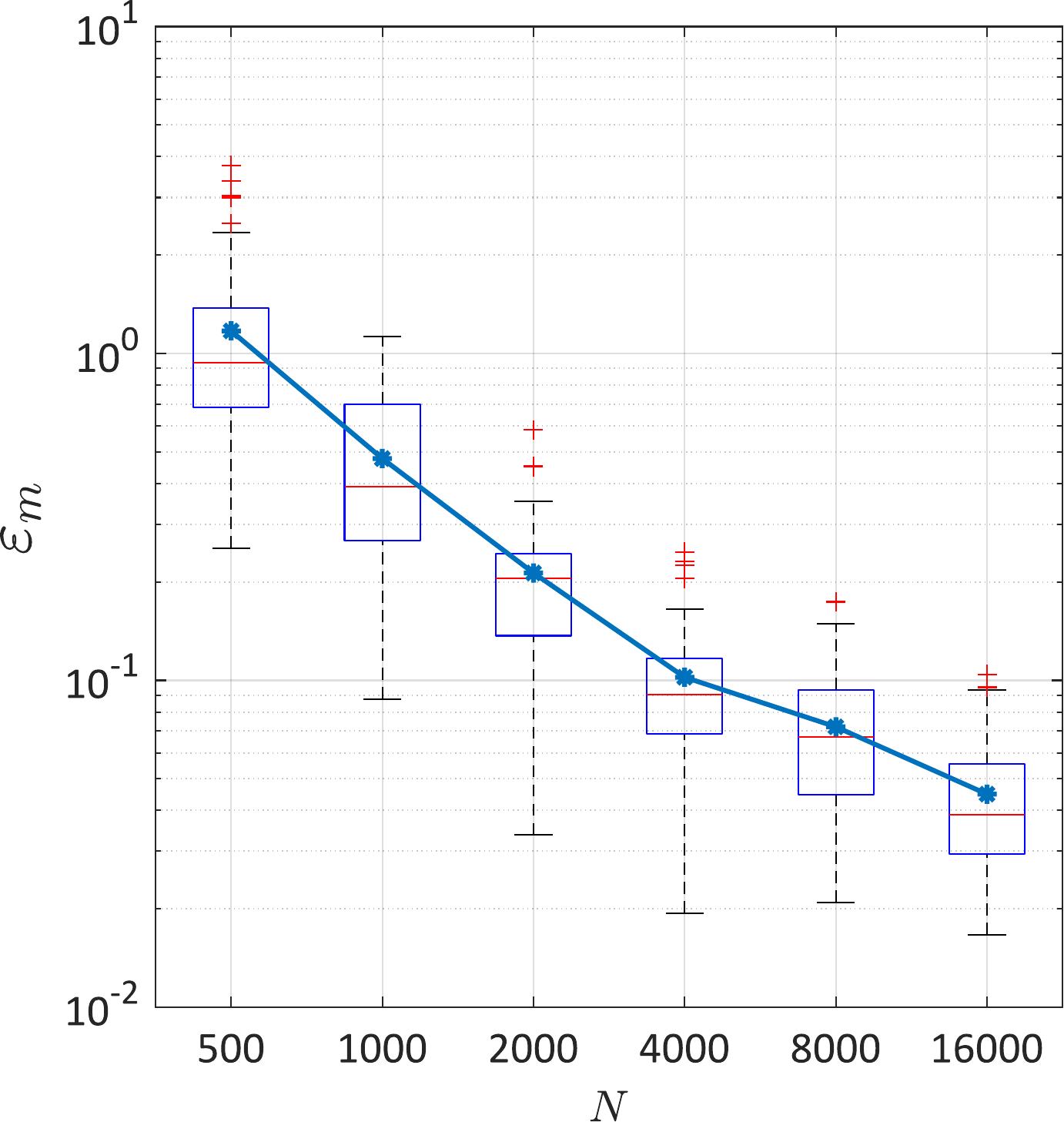}
		\caption{Mean estimation}
	\end{subfigure}
	\hspace{0.5cm}
	\begin{subfigure}[!b]{.45\linewidth}
		\centering
		\includegraphics[height=1\linewidth, 
		keepaspectratio]{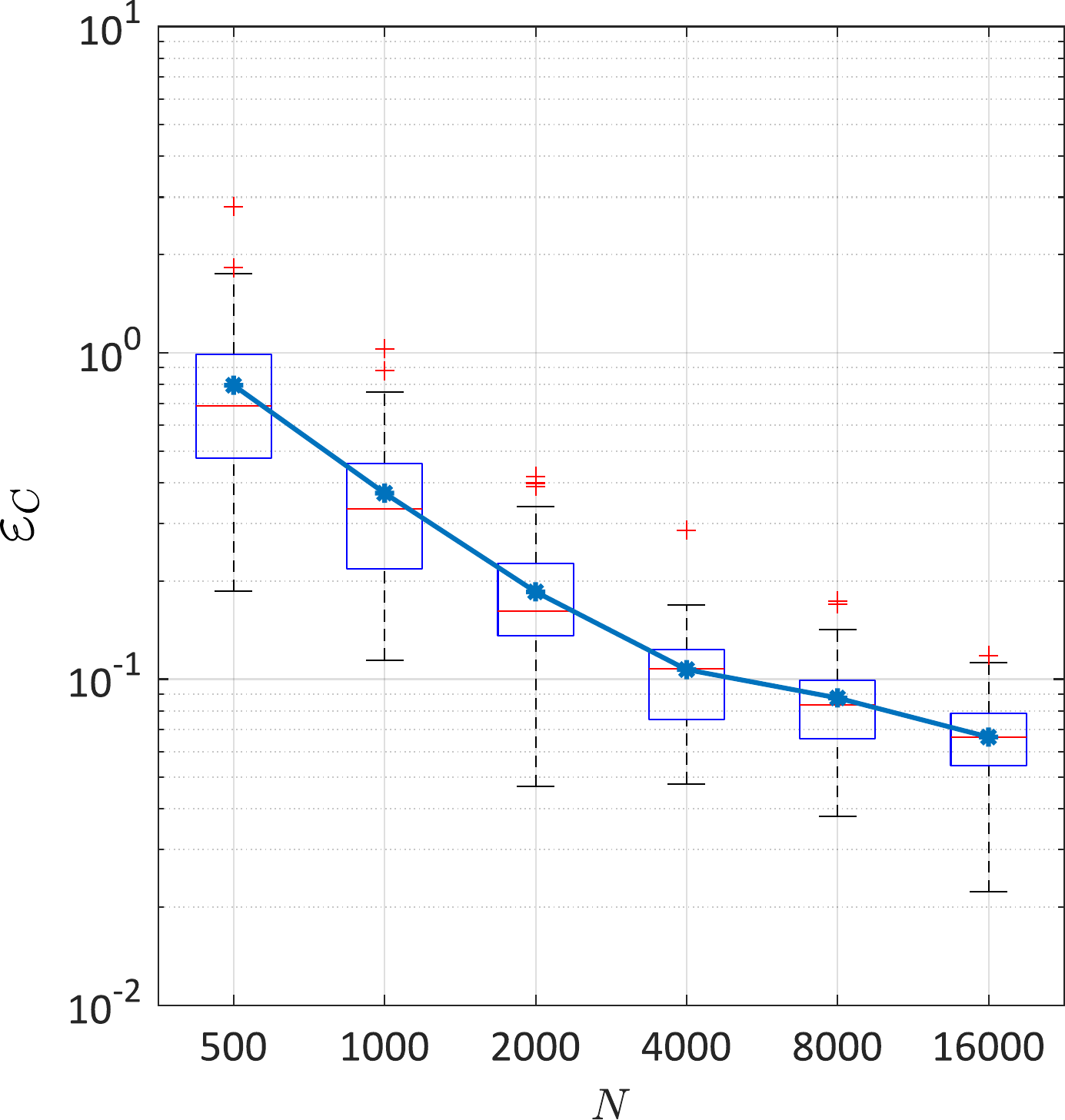}
		\caption{Expected payoff estimation}
		\label{fig:Hestonconvp}
	\end{subfigure}	
	\caption{Heston model -- Convergence study. The blue lines denote the 
		errors averaged over 50 repetitions of the full analysis.}
	\label{fig:Hestonconv}
\end{figure}

\Cref{fig:Hestonconv} shows box plots of the errors $\varepsilon_{m}$ 
and $\varepsilon_{c}$ for $N\in 
\{500 ; 1{,}000 ; \allowbreak 2{,}000 ; \allowbreak 4{,}000 ; \allowbreak 
8{,}000 ; 16{,}000\}$. Both $\varepsilon_{m}$ and $\varepsilon_{c}$ are 
relatively large for $N \leq 2{,}000$. This is 
mainly due to the fact that the 
variability of the model 
response is dominated by the intrinsic randomness: the model parameters 
$\ve{X}$ altogether are only able to explain about 2\% of the variance of the 
output (\ie $S_{\acc{1,\ldots,6}} = 0.02$). In 
other words, the stochastic simulator has a very small signal-to-noise ratio 
$\SNR = 0.02/(1-0.02) \approx 0.02$. Since GLDs are flexible, a few data 
scattered in a moderately high dimensional space may not provide enough 
information of the response distribution variation. We observe that for 
$N\leq 1{,}000$, the selection procedure 
proposed in \Cref{alg:FGLS} can 
choose 
$\lambda^{\PC}_1(\ve{x})$ and $\lambda^{\PC}_2(\ve{x})$ being only constant. 
Such a model is too simple and thus fails to capture the variations 
of the scalar quantities. Consequently, it is necessary to have enough data to 
achieve an accurate estimate: when increasing the size of $N$ of the LHS 
design, we observe a clear decay of the 
errors. 

\par
\par
We now study the convergence for the Sobol' indices estimations. According to 
\Cref{eq:HestonMean}, the mean function depends only on the first input 
variable $X_1$, which contributes little (2\%) to the total variance of 
$Y_1(\ve{X})$. 
This implies that the classical Sobol' indices are not informative (they 
are either 0 or very close to 0). However, we cannot 
ignore the variability of the input variables because the response 
distribution demonstrates a clear dependence on the input parameters, as shown 
in \Cref{fig:Hestonpdf}. Therefore, we focus on the accuracy of the 
expected-payoff-based total Sobol' indices, denoted by $S^C_{T_{\iu}}$. 

As a second quantity of interest, we also calculate the total Sobol' indices 
associated to the $95\%$-superquantile, referred to as $S^{\rm sq}_{T_{\iu}}$. 
Superquantiles are known as the \emph{conditional value-at-risk}, which is 
an important risk measure in finance \cite{Acerbi2002}. The 
$\alpha$-superquantile of a random variable $Y$ is defined by
\begin{equation}\label{eq:CVaR}
{\rm sq}_{\alpha} = \Esp{Y \mid Y \geq q_{\alpha}},
\end{equation}
where $q_{\alpha}$ is the $\alpha$-quantile of $Y$. This quantity corresponds to
the conditional expectation of $Y$ being larger than its $\alpha$-quantile. 
For the Heston model, this quantity does not have an analytical closed form, 
whereas ${\rm sq}_{\alpha}$ of a GLD can be derived analytically (see 
\Cref{sec:pGLDs}).

We use $10^5$ Monte Carlo samples to evaluate (numerically) the function 
$m_C(\ve{x})$ to obtain a reference value for each $S^C_{T_{\iu}}$. To 
calculate the Sobol' indices associated to the 95\%-superquantile ${\rm 
sq}_{95}(\ve{X})$, it is necessary to evaluate the function ${\rm 
sq}_{95}(\ve{x})$. Because it cannot be analytically derived for the Heston 
model, we use $10^4$ replications to calculate the sample 95\%-superquantile 
$\hat{{\rm sq}}_{95}(\ve{x})$ as an estimate for ${\rm sq}_{95}(\ve{x})$. Then, 
we treat it as a deterministic function and use $10^4$ samples to estimate 
each Sobol' index. This indicates that a total number of $7 \times 10^8$ model 
runs are performed to obtain the six reference 95\%-superquantile-based 
total Sobol' indices. Because only $10^4$ samples are used to estimate each 
$S^{\rm sq}_{T_{\iu}}$, we use bootstraps \cite{Efron1979} to calculate the 
95\% confidence interval to account for the uncertainty of the Monte Carlo 
simulation. 

\begin{figure}[!htbp]
	\centering	
	\includegraphics[width=0.98\linewidth, 
	keepaspectratio]{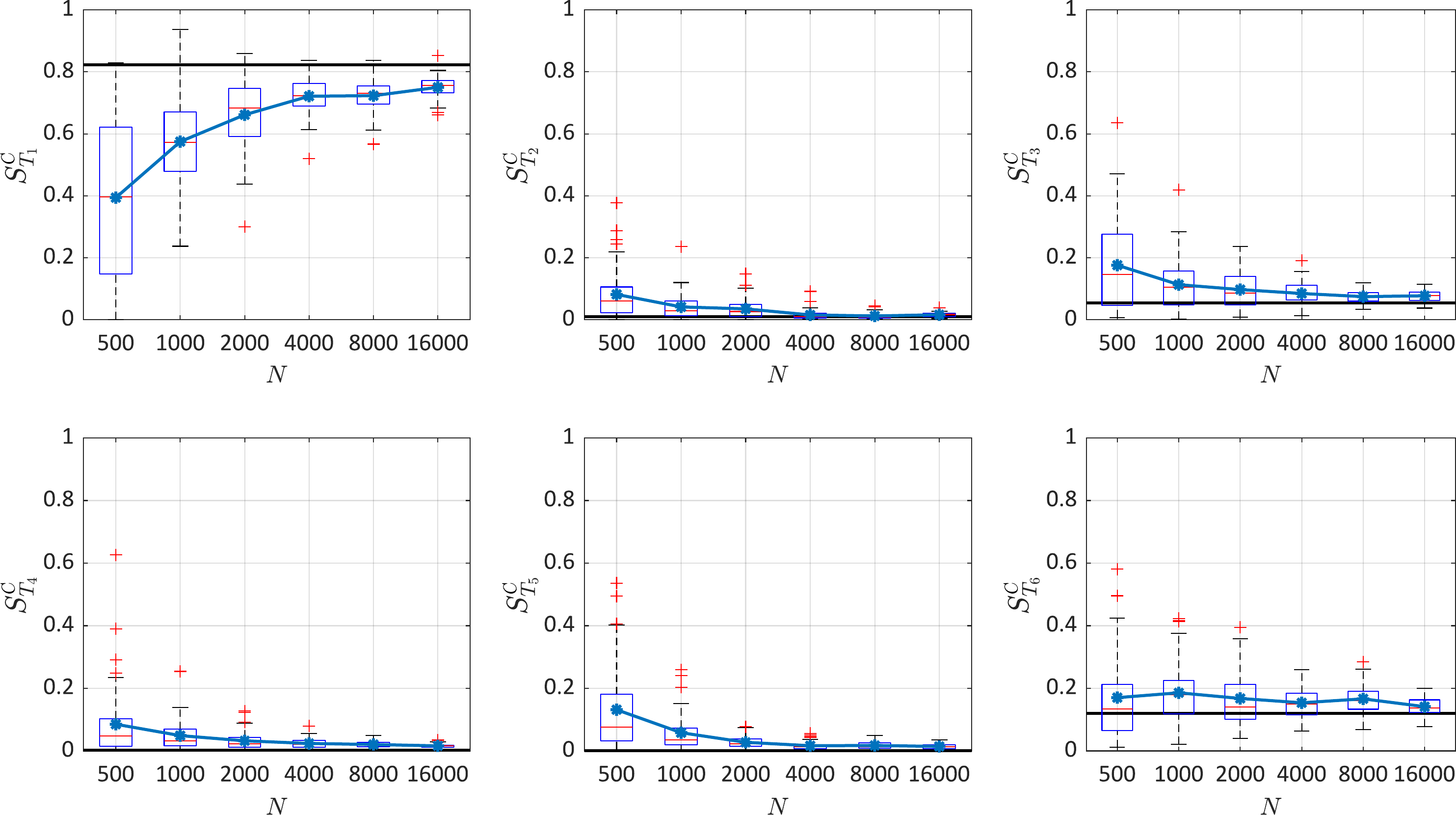}
	\caption{Estimation of the expected-payoff-based total Sobol' indices. The 
	black lines are the reference values, and the blue lines denote the 
	average values of 50 repetitions.}
	\label{fig:Hestonsen1}
\end{figure}

\begin{figure}[!htbp]
	\centering	
	\includegraphics[width=0.98\linewidth, 
	keepaspectratio]{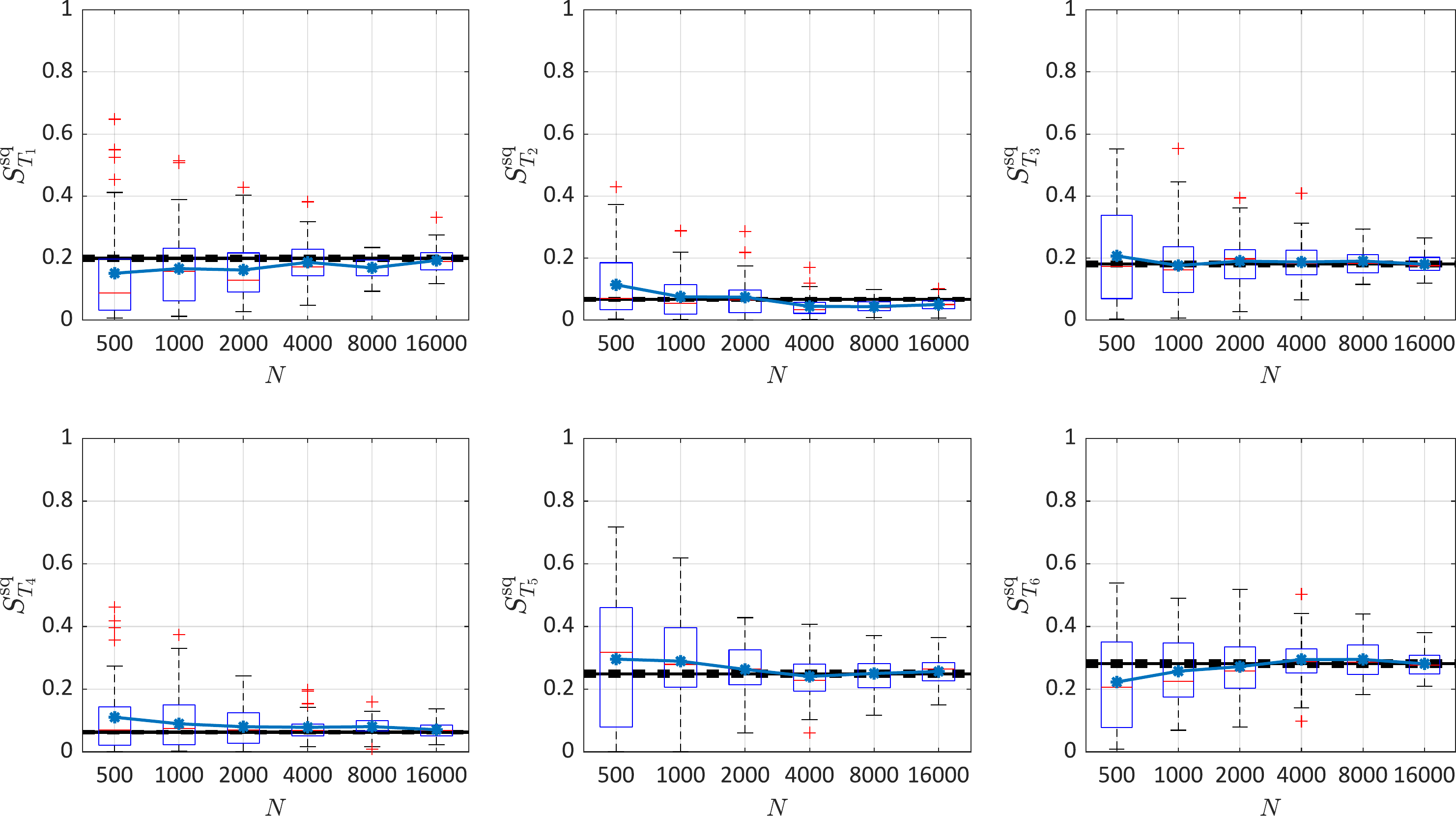}
	\caption{Estimation of the 95\%-superquantile-based total Sobol' indices. 
	The blue lines denote the average value of 50 repetitions of the full 
	analysis. The black lines are the reference values, and the dashed lines 
	correspond to the 95\% confidence intervals.}
	\label{fig:Hestonsen2}
\end{figure}

\Cref{fig:Hestonsen1,fig:Hestonsen2} confirm and quantify the 
convergence of GLaMs to estimate $S^{C}_{T_{\iu}}$ and $S^{\rm sq}_{T_{\iu}}$.
For the expected payoff $m_{C}(\ve{x})$, the first variable $\mu$ is the most 
important. The estimation of its total effect converges from below the 
reference value, and we observe a bias in the estimate. Nevertheless, with $N$ 
large enough ($\geq 4{,}000$), the GLaM can 
always correctly identify its importance (the bias is 0.072 for $\geq 8{,}000$ 
and 0.055 for $\geq 16{,}000$), and each classical first-order Sobol' index of 
the 
other five variables converge to the reference line. The 95\%-superquantile 
suggests a different ranking: $\mu$, $\theta$, $\rho$ and $v_0$ (corresponding 
to the first, third, fifth and sixth input variable, respectively) have similar 
total effects, which are superior to those of $\kappa$ and $\sigma$ (\ie the 
second and fourth input variables). In addition, none of the input variables 
has nearly 0 total effect. The GLaM surrogate model accurately reproduces the 
phenomena. Moreover, the estimates generally vary around the reference values, 
and larger $N$ results in narrower spread of the box plots. 

As a conclusion, GLaM surrogates allow us to represent accurately the QoI of 
the Heston model and carry out a detailed sensitivity analysis at the cost of 
$\mathcal{O}(10^4)$ runs of the stochastic simulator. Note that in this example 
the leave-one-out errors of the polynomial chaos expansions built on the GLaM 
surrogates are of the order of $o(10^{-4})$, which justifies the use of 
PCE-based Sobol' indices.

\subsection{Stochastic SIR model}
In this example, we apply the proposed method to a \emph{stochastic 
Susceptible-Infected-Recovered} (SIR) model in epidemiology \cite{Britton2010}. 
This model simulates the spread of an infectious disease, which can help 
conduct appropriate epidemiological intervention to minimize the social and 
ethical impacts during the outbreak.
\par
In a SIR model, a population of size $P_t$ at time $t$
can be partitioned into three groups: susceptible, infected and recovered 
during the outbreak of an epidemic. Susceptible individuals are those who can 
get infected by contacting an infectious person. Infected individuals are 
suffering from the disease and are contagious. They can recover (therefore 
classified as recovered) and become immune to future infections. The number of 
individuals within each group is denoted by $E_t$, $I_t$ and $R_t$, 
respectively. Without differentiating individuals, these three quantities 
characterize the configuration of the population at a given time $t$. Hence, 
their evolutions represent the spread of the epidemic. In this study, we 
consider a fixed population without newborns and deaths, \ie the total 
population size is constant, $P_t = P$. As a result, $E_t$, $I_t$ and $R_t$ 
satisfy the constraint $E_t+I_t+R_t = P$, and thus only the time evolution of 
$E_t$ and $I_t$ is necessary to represent the disease evolution. 

\begin{figure}[!htbp]
	\centering
	\includegraphics[width=0.7\linewidth]{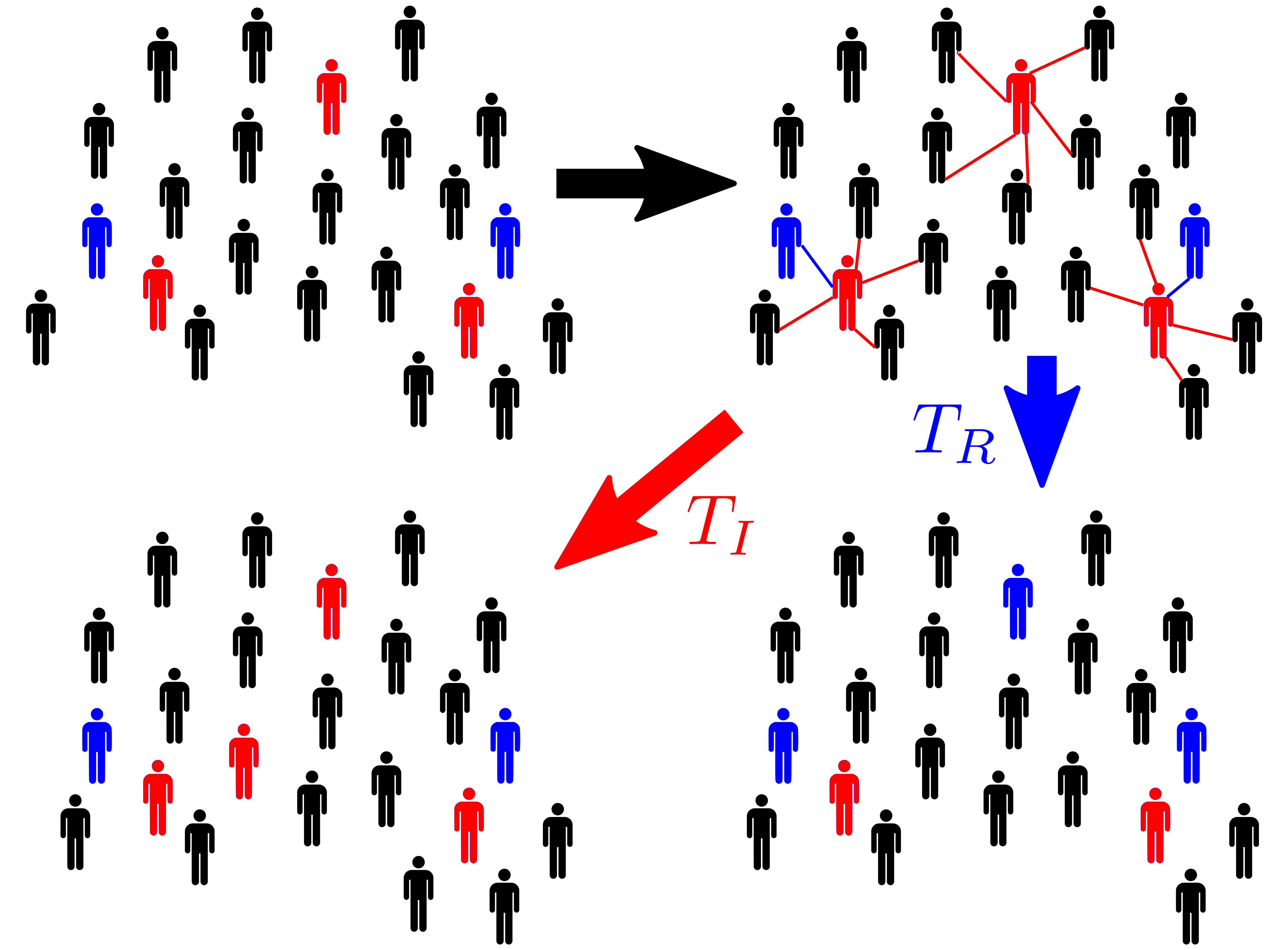}
	\caption{Dynamics of the stochastic SIR model: black icons denote 
	susceptible individuals, red icons represent infected individuals, and blue 
	icons are those recovered.}
	\label{fig:SIR}
\end{figure}

Without going into detailed assumptions of the model, we illustrate 
the system dynamics in \Cref{fig:SIR}, where the black icons represent 
susceptible individuals, the red icons indicate infected persons, and the blue 
icons are those recovered. Suppose that at time $t$ the population has the 
configuration $(E_t,I_t)$ (top left figure of \Cref{fig:SIR}). Infected 
individuals can meet susceptible individuals, or they may receive essential 
treatments and recover from the disease. Hence, the next configuration has two 
possibilities: (1) $C_I$, where one susceptible individual is infected; (2) 
$C_R$, where one infected person recovers. The population state evolving 
either to $C_I$ or $C_R$ depends on two random variables, $T_I$ and $T_R$, 
which denote the respective time to move to the associated candidate 
configuration. Both random variables follow an exponential distribution, yet 
with different parameters: 
\begin{align}
T_I &\sim \Exp(\lambda_I), \quad \lambda_I = \beta \frac{E_t I_t}{P}, \\
T_R &\sim \Exp(\lambda_R), \quad \lambda_R = \gamma I_t,
\end{align}
where $\beta$ indicates the contact rate of an infected individual, and 
$\gamma$ is the recovery rate. If $T_I>T_R$, $C_I$ becomes the next 
configuration at $t+T_I$ with $S_{t+T_I}=E_t-1$ and $I_{t+T_I}=I_t+1$, and vice 
versa. This update step iterates until time $T$ when $I_T = 0$. Because the 
population size is finite and the recovered individuals will not get infected 
again, the total number of updates is finite ($\leq P$). This number is not a
constant due to the updating process, indicating that the amount of latent 
variables of this simulator is also random. Note that the evolution 
procedure described here corresponds to the \emph{Gillespie algorithm} 
\cite{Gillespie1977}. 

In this case study, we set $P=2{,}000$. $\ve{x}=(E_0,I_0,\beta,\gamma)$ is the 
vector of input parameters. To account for different scenarios, the input 
variables $\ve{X}$ are modeled as $X_1 \sim \cu(1{,}600\,,\,1{,}800)$, $X_2 \sim 
\cu(20,200)$ and $X_3, X_4 \sim \cu(0.5,0.7)$. The uncertainty in the first two 
variables can be interpreted as lack of knowledge of the initial condition. 
While the last two variables are affected by possible interventions, 
such as social distancing measures that can reduce the contact rate $\beta$ and 
increasing medical resources that improves the recovery rate $\gamma$. We are 
interested in the total number of newly infected individuals during 
the outbreak, \ie $E_T - E_0$. The signal-to-noise ratio of this stochastic 
model is estimated to be $\SNR \approx 6.7$, which is relatively large.

\begin{figure}[!htbp]
	\centering
	\begin{subfigure}[!b]{.45\linewidth}
		\centering
		\includegraphics[height=0.8\linewidth, keepaspectratio]{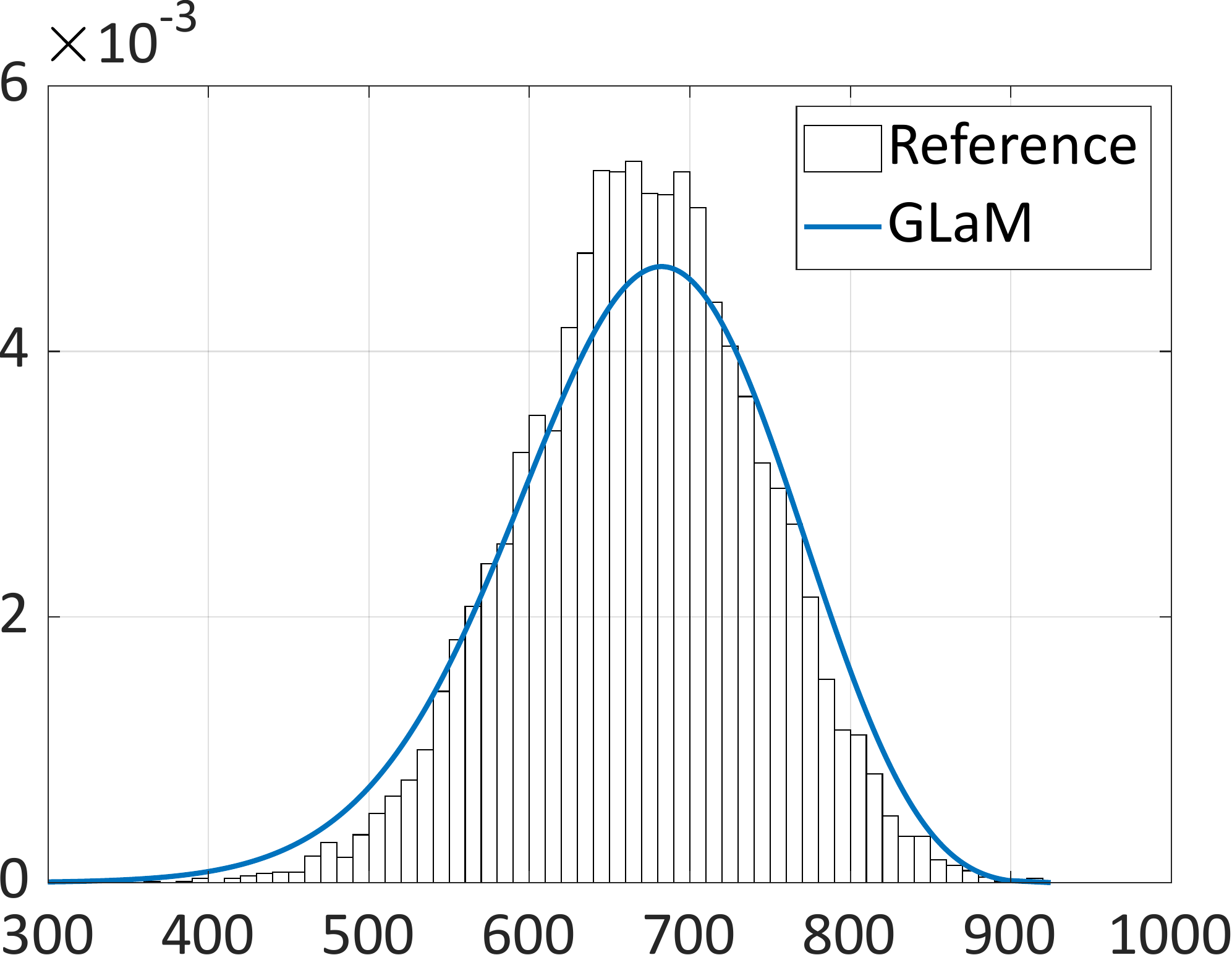}
		\caption{PDF for $\ve{x} = (1750,150,0.55,0.65)$}
		\label{subfig:sirpdf1}
	\end{subfigure}
	\hspace{0.5cm}
	\begin{subfigure}[!b]{.45\linewidth}
		\centering
		\includegraphics[height=0.8\linewidth, keepaspectratio]{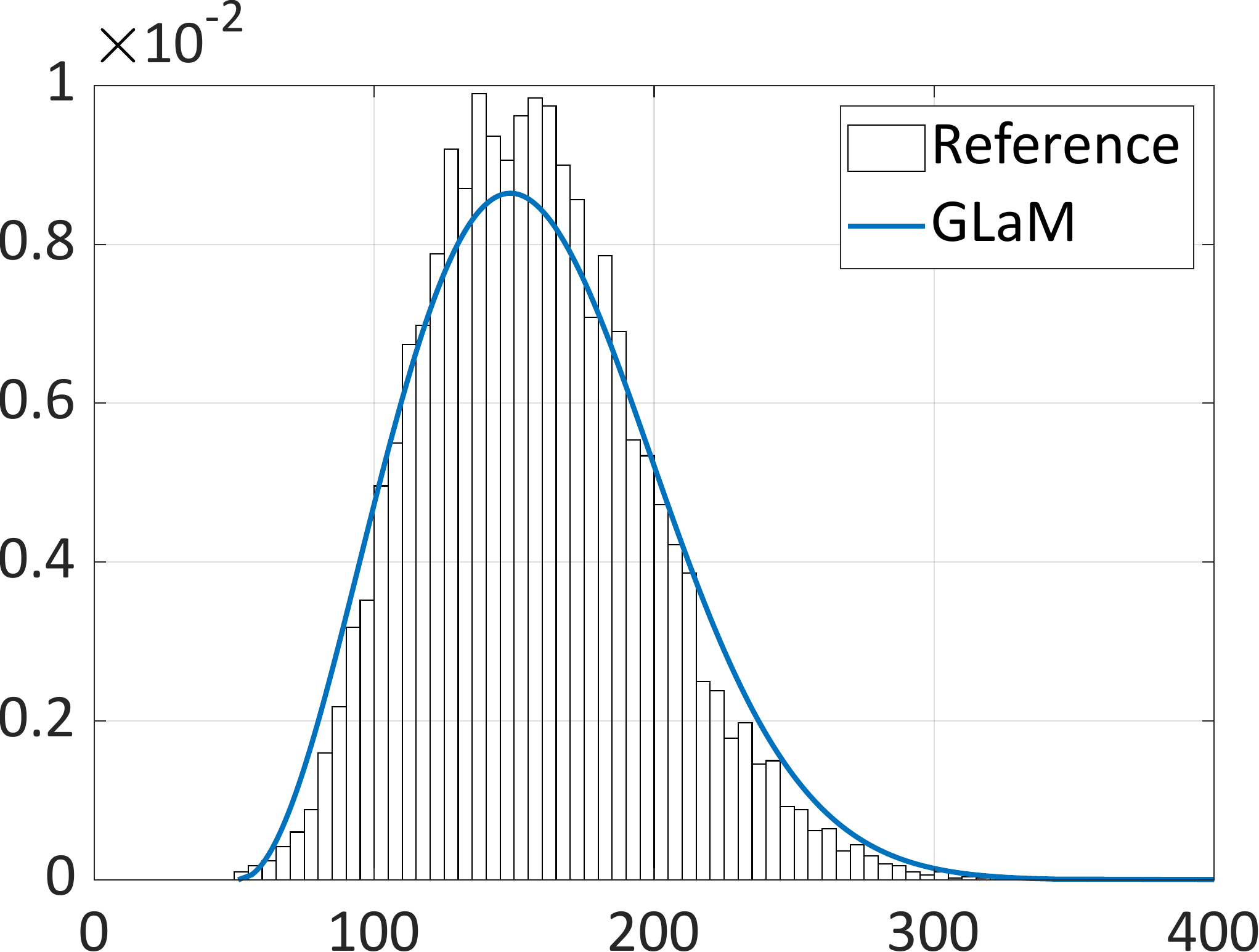}
		\caption{PDF for $\ve{x} = (1650,80,0.65,0.55)$}
		\label{subfig:sirpdf2}
	\end{subfigure}
	\caption{Stochastic SIR model -- Emulated response PDFs, $N=1{,}000$}
	\label{fig:sirpdf}
\end{figure}

\Cref{fig:sirpdf} shows the response PDF estimation of the surrogate model 
built on an experimental design of $N=1{,}000$. The reference histograms are 
calculated from $10^4$ repeated model runs with the same input values. We 
observe that the response distribution changes from right-skewed to left-skewed 
distributions (so we can also find symmetric distributions in between), which 
is correctly represented by the surrogate. In addition, GLaMs also accurately 
approximate the bulk and the support of the response PDF. 

In this example, we investigate the convergence of GLaMs for estimating the 
classical first-order Sobol' indices and the standard-deviation-based total 
Sobol' indices, denoted by $S^{\sigma}_{T_{\iu}}$. To calculate the reference 
values, we use $10^5$ Monte Carlo 
samples for each classical Sobol' index. Regarding the 
standard-deviation-based Sobol' indices, we calculate the sample standard 
deviation $\hat{\sigma}(\ve{x})$ based on $10^4$ replications. 
Then, we apply Monte Carlo simulations with $10^4$ samples to estimate the 
associated Sobol' indices. The total cost to get reference values is thus equal 
to $5\times 10^8$. As in the previous example, we use bootstraps to 
calculate the 95\% confidence intervals. 

\begin{figure}[!htbp]
	\centering
	\includegraphics[width=0.95\linewidth, 
	keepaspectratio]{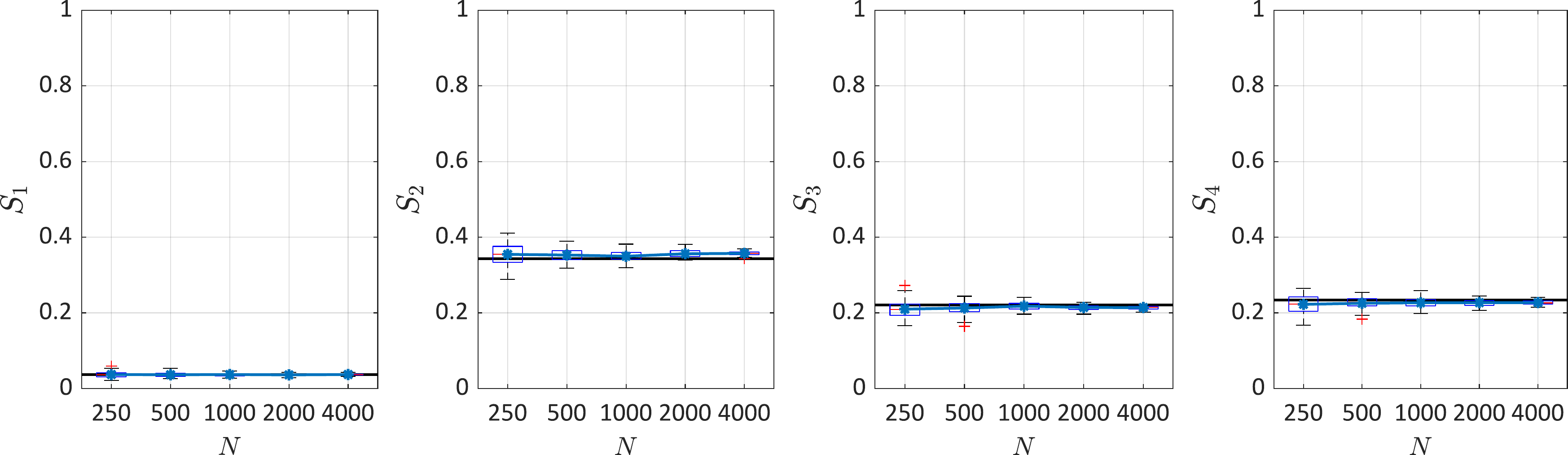}
	\caption{Estimation of the classical first-order Sobol' indices. The black 
		lines are the reference values, and the blue lines denote the average 
		values of 50 repetitions of the full analysis.}
	\label{fig:SIRsen1}
\end{figure}

\begin{figure}[!htbp]
	\centering	
	\includegraphics[width=0.95\linewidth, 
	keepaspectratio]{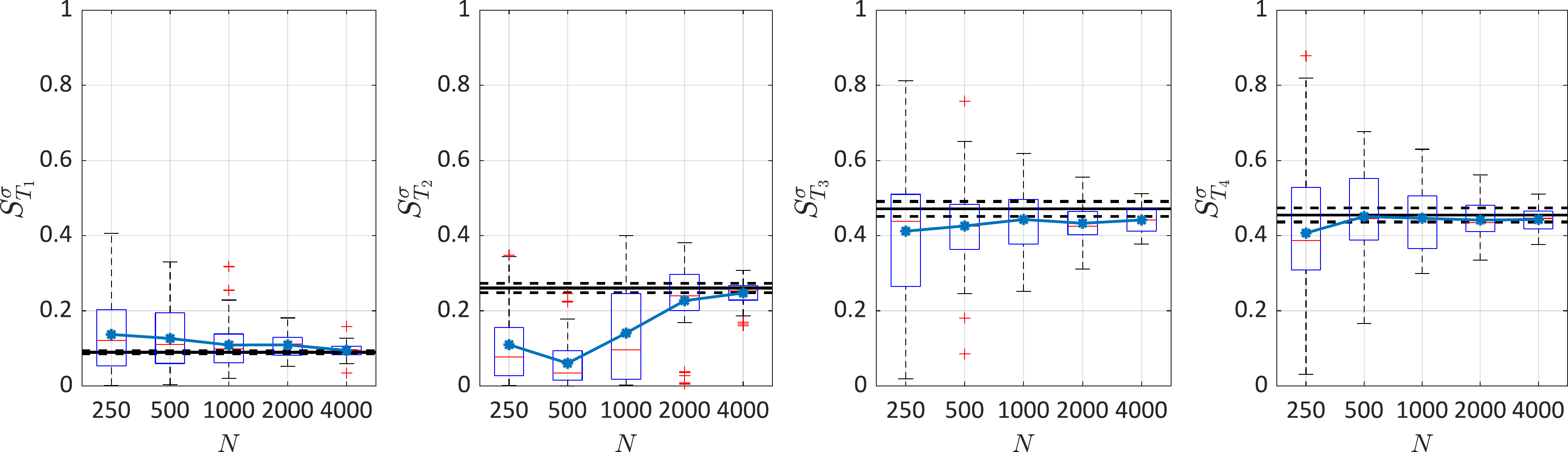}
	\caption{Estimation of the standard-deviation-based total Sobol' indices. 
	The blue lines denote the average values of 50 repetitions of the full 
	analysis. The black 
	lines are the reference values, and the dashed lines correspond to the 95\% 
	confidence intervals.}
	\label{fig:SIRsen2}
\end{figure}

\Cref{fig:SIRsen1,fig:SIRsen2} show the results of the convergence 
study. In terms of the classical Sobol' indices, the GLaM yields 
accurate estimates even when $N=250$: the box plots scatter around the 
reference values with a small variability. Among the four input variables, the 
second one $I_0$ that corresponds to the number of infected individuals at time 
0 is the most important. It is followed by the contact rate and the recovery 
rate, which show similar first-order effect. As a result, performing medical 
test to better determine $I_0$ would be the most effective way to reduce the 
variance of the output. In contrast, \Cref{fig:SIRsen2} suggests that 
controlling the contact rate and recovery rate would be the best measure to 
reduce the variation of $\sigma(\ve{X})$. For estimating the 
associated Sobol' indices, the GLaM converges within the 95\% confidence 
intervals of the Monte Carlo estimates, and the spread of the box plots 
decreases significantly with $N$ increasing.

Finally, we remark that when higher-order Sobol' indices are of interest, 
Monte Carlo simulations require additional runs of the original model. For 
example, $4 \times 10^8$ more model evaluations should be performed 
to obtain the reference values for the standard-deviation-based second-order 
Sobol' indices. This results in a large amount of total model runs, which 
become impracticable even with cheap models. In contrast, GLaM surrogates can be 
used without additional cost: the PCE-based method presented in 
\Cref{sec:SGLaM3} provides analytical higher-order indices by post-processing 
the PC coefficients. In this example, the leave-one-out errors of PCE built on 
the GLaM surrogates are of the order $\mathcal{O}(10^{-3})$, which justifies 
the use of PCE estimates. Based on the surrogate model, we observe that the 
largest standard-deviation-based second-order Sobol' indices, which translate 
parameters interactions, are $I_0$ and $\beta$, $I_0$ and $\gamma$. Both have a 
value 0.09, while the other second-order interactions are very small. As 
illustrated in \Cref{fig:SIRsen2}, $I_0$ has a total effect $S^{\sigma}_{T_2} 
=0.26$. Moreover, it has a relatively small first-order effect $S^{\sigma}_2 
= 0.05$. This implies that $I_0$ mainly affects the variance of 
$\sigma(\ve{X})$ through its interactions with $\beta$ and $\gamma$.

\section{Conclusions}
\label{sec:conclusions}

In this paper, we discuss the nature and focus of three extensions of Sobol' 
indices to stochastic simulators: classical Sobol' indices, QoI-based Sobol' 
indices and trajectory-based Sobol' indices. The first two types are of 
interest because of their versatility and applicability to a broad class of 
problems. We propose to use the generalized lambda model as a stochastic 
emulator to estimate the considered indices. This surrogate model aims at 
emulating the entire response distribution, instead of focusing only on some 
scalar statistical quantities, \eg mean and variance. More precisely, it
relies on using the four-parameter generalized lambda distribution to 
approximate the response distribution. The associated distribution parameters 
as functions of the input are represented by polynomial chaos expansions. Such 
a surrogate can be constructed without the need for replications, and thus it 
is not restricted to a special data structure. 

Because of the special formulation of GLaM, the considered sensitivity indices 
can be estimated by directly working with deterministic functions. This allows 
applying the methods developed for deterministic simulators, namely estimators 
based on Monte Carlo simulations and on polynomial chaos expansions. In this 
paper, we suggest the latter to post-process the surrogate model to achieve 
high computational efficiency.

The performance of the proposed method for estimating various Sobol' indices 
is illustrated on three examples with different signal-to-noise ratios. The 
toy example is designed to have a strong heteroskedastic effect. It shows the 
general convergent behavior of GLaMs for approximating the conditional quantile 
functions and estimating the entropy of the response distributions. The second 
example is a Heston model from mathematical Finance. This case study has a very 
small signal-to-noise ratio and 
demonstrates a shape variation of the response PDF. The surrogate generally 
yields accurate estimate of the Sobol' indices associated to the expected 
payoff and the 95\%-superquantile. The last example is a stochastic SIR model 
in epidemiology, in which GLaMs exhibit robust estimates of the classical 
Sobol' indices and the standard-deviation-based Sobol' indices. All three 
examples have a different ranking of the input variables depending on the type 
of Sobol' indices, which is correctly captured by GLaMs when comparing to 
reference values obtained by extremely costly Monte Carlo simulations. Fairly 
accurate results are obtained at a cost of $\mathcal{O}\left(10^4\right)$ runs 
of the simulator compared to reference values based on 
$\mathcal{O}\left(10^8\right)$ runs by a brute force approach.

In future work, we plan to develop algorithms to improve GLaMs for small 
data sets. Besides, we will investigate GLaMs for estimating distribution-based 
sensitivity indices \cite{Borgonovo2007,Huoh2013}. The estimation of these 
indices usually requires a large number of model runs to infer the 
conditional PDF, which can be easily obtained from GLaMs. In addition, 
appropriate contrast measures between distributions, such as
the Wasserstein metric, can be developed for sensitivity analysis in the 
context of stochastic simulators. Finally, developing sensitivity 
indices for stochastic simulators with dependent input variables will allow 
engineers to tackle a broader group of problems.

\section*{Acknowledgements}
This paper is a part of the project ``Surrogate Modeling for Stochastic 
Simulators (SAMOS)'' funded by the Swiss National Science Foundation (Grant 
\#200021\_175524), the support of which is gratefully acknowledged.

%\section*{References}
\section{Appendix}
\subsection{Some properties of GLDs}
\label{sec:pGLDs}
The mean and variance of a GLD can be calculated by
\begin{align}
m & = \lambda_1 - 
\frac{1}{\lambda_2}\left(\frac{1}{\lambda_3+1} 
- \frac{1}{\lambda_4(\ve{x})+1}\right), \label{eq:GLD_mean}\\
v & = \frac{(d_2-d_1^2)}{\lambda_2^2}, 
\label{eq:GLD_var}
\end{align}
where $\acc{d_k: k=1,2}$ are defined by
\begin{equation}\label{eq:GLD_MM_aux}
\begin{split}
d_1 &= \frac{1}{\lambda_3}\Betafun(\lambda_3+1,1) - 
\frac{1}{\lambda_4}\Betafun(1,\lambda_4+1), \\
d_2 &= \frac{1}{\lambda^2_3}\Betafun(2\lambda_3+1,1) - 
\frac{2}{\lambda_3\lambda_4}\Betafun(\lambda_3+1,\lambda_4+1) +
\frac{1}{\lambda^2_4}\Betafun(1,2\lambda_4+1),
\end{split}
\end{equation}
with $\Betafun$ denoting the beta function. 

The expected payoff defined in \Cref{eq:epoff} of a GLD with the strike price 
$K$ is given by 
\begin{equation}\label{eq:epoff}
\begin{split}
m_C &\eqdef \Esp{\max\acc{Y-K,0}} \\
&= \left(\lambda_1 - \frac{1}{\lambda_2 \lambda_3} + \frac{1}{\lambda_2 
\lambda_4} - K\right)\,(1-u_K) + 
\frac{1}{\lambda_2}\,\left(
\frac{1-u_K^{\lambda_3+1}}{\lambda_3\,\left(\lambda_3+1\right)}
-\frac{(1-u_K)^{\lambda_4+1}}{\lambda_4\,\left(\lambda_4+1\right)}
\right).
\end{split}
\end{equation}
where $u_K$ is the solution of the nonlinear equation:
\begin{equation}
Q(u_K;\ve{\lambda}) = K.
\end{equation}

The $\alpha$-superquantile ${\rm sq}_{\alpha}$ defined in \Cref{eq:CVaR} of a 
GLD has a closed-form:
\begin{equation}\label{eq:supq}
\begin{split}
{\rm sq}_{\alpha} &\eqdef \Esp{Y \mid Y> q_{\alpha}} \\
&= \lambda_1 - \frac{1}{\lambda_2 \lambda_3} + \frac{1}{\lambda_2 
	\lambda_4} + 
\frac{1}{(1-\alpha)\lambda_2}\,\left(
\frac{1-\alpha^{\lambda_3+1}}{\lambda_3\,\left(\lambda_3+1\right)}
-\frac{\alpha^{\lambda_4+1}}{\lambda_4\,\left(\lambda_4+1\right)}
\right).
\end{split}
\end{equation}
\bibliography{mybibfile}

\end{document}